\documentclass[iop]{emulateapj}
\usepackage{amsmath}
\usepackage{threeparttable}
\usepackage{graphicx}
\usepackage{rotating}
\usepackage{subfloat}
\usepackage{color}
\usepackage{natbib}
\usepackage{soul}
\usepackage{subfigure}
\usepackage{longtable}              
\usepackage{setspace}
\setcitestyle{authoryear,round,semicolon,aysep={,},yysep={,},notesep={; }}
\begin{document}
\title{Inferred H$\alpha$ Flux as a star-formation rate
  indicator at $\lowercase{z}\sim4$-5: Implications for dust properties,
  burstiness, and the $\lowercase{z}=4$-8 Star-Formation-Rate
  Functions} 
 \author{Renske Smit\altaffilmark{1,2}, Rychard J. Bouwens\altaffilmark{2}, 
 Ivo Labb\'{e}\altaffilmark{2}, Marijn Franx\altaffilmark{2},
  Stephen M. Wilkins\altaffilmark{3}, Pascal A. Oesch\altaffilmark{4}
 }
 \altaffiltext{1}{Centre for Extragalactic Astronomy, Durham University, South Road, Durham, DH1 3LE, UK} 
\altaffiltext{2}{Leiden Observatory, Leiden University, NL-2300 RA Leiden, Netherlands} 
\altaffiltext{3}{Astronomy Centre, Department of Physics and Astronomy, University of Sussex, Brighton BN1 9QH, UK}
\altaffiltext{4}{Yale Center for Astronomy and Astrophysics, Yale University, New Haven, CT 06520, USA}

\begin{abstract}
We derive H$\alpha$ fluxes for a large spectroscopic and
photometric-redshift-selected sample of sources over GOODS-North and
South in the redshift range $z=3.8$-5.0 with deep \text{HST},
\textit{Spitzer}/IRAC, and ground-based observations.  The H$\alpha$
flux is inferred based on the offset between the IRAC $3.6\,\micron$
flux and that predicted from the best-fit SED.  We demonstrate that
the H$\alpha$ flux correlates well with dust-corrected UV
star-formation rate (SFR) and therefore can serve as an independent
SFR indicator.  However, we also find a systematic offset in the $\rm
SFR_{\rm H\alpha}/SFR_{\rm UV+\beta}$ ratios for $z\sim4$-5 galaxies
relative to local relations (assuming the same dust corrections for
nebular regions and stellar light).
We show that we can resolve the modest tension in the inferred SFRs by
assuming bluer intrinsic UV slopes (increasing the dust correction), a
rising star-formation history or assuming a low metallicity stellar
population with a hard ionizing spectrum (increasing the $\rm
\textit{L}_{\rm H\alpha}/SFR$ ratio).  Using H$\alpha$ as a SFR indicator, we find a normalization of the star
formation main sequence in good agreement with recent SED-based determinations
and also derive the SFR functions at $z\sim4$-8.  In addition, we
assess for the first time the burstiness of star formation in $z\sim4$
galaxies on $<$100 Myr time scales by comparing $UV$ and
H$\alpha$-based sSFRs; their one-to-one relationship argues against
significantly bursty star-formation histories.  Further progress will
be made on these results, by incorporating new results from ALMA to
constrain the dust-obscured star formation in high-redshift
UV-selected samples.

\end{abstract}

\section{Introduction}
\label{sec:intro}

Over the last decade, dedicated deep field programs with the \textit{Hubble Space Telescope (HST)} have identified more than 10000 candidate galaxies with a redshift beyond $z\gtrsim4$, based on their photometric colors \citep[e.g.][]{Bouwens2014}. Although a number of these objects have been successfully confirmed out to $z\sim8.7$ through near-infrared (NIR) spectroscopy \citep[e.g.][]{Zitrin2015,Oesch2015,Finkelstein2013}, progress in characterizing the spectral energy distributions (SEDs) of these galaxies and identifying their physical properties has been slow. This is largely due to the fact that spectroscopy and deep, high-resolution photometry in the rest-frame optical wavelengths, shifted to observed mid-infrared (MIR) wavelengths for sources at $z\gtrsim4$, will not be available until the launch of the \textit{James Webb Space Telescope (JWST)}. 

Despite these challenges a number of noteworthy results have emerged on the observational properties of the $z\gtrsim4$ galaxy population. First of all, the typical rest-frame UV colors of galaxies from $z\sim4$ to $z\sim8$ have been meticulously characterized through their \textit{HST} photometry \citep{Bouwens2009, Bouwens2012,Bouwens2013,Dunlop2012,Dunlop2013,Wilkins2011,Finkelstein2012}. Furthermore, photometric studies using the \textit{Spitzer Space Telescope} have obtained the first constraints on the shape of the rest-frame optical SED of $z\gtrsim4$ galaxies \citep{Eyles2005,Verma2007,Wilklind2008,Yabe2009,Stark2009,Labbe2010a,Labbe2010b,Gonzalez2010,Gonzalez2012}. In particular, observational evidence has emerged for the presence of strong optical nebular emission lines, such as H$\alpha$ and [\ion{O}{3}], in the typical $z\gtrsim4$ source \citep{Schaerer2009,Shim2011,Stark2013,Labbe2013,Gonzalez2014,deBarros2014,Smit2014,Smit2015,Roberts2015,Rasappu2015,Marmol2015}. 
 
While the rest-frame equivalent widths (EWs) of H$\alpha$ in typical star-forming galaxies at $z\sim0$-2 range from 10-200{\AA} \citep{Fumagalli2012}, a large fraction of sources between $z\sim4$-8 have inferred H$\alpha$ and [\ion{O}{3}] EWs in the range $250-600${\AA} and $600-1000${\AA} respectively \citep{Schaerer2009,Shim2011,Stark2013,Labbe2013,Gonzalez2014,deBarros2014,Smit2014,Smit2015,Rasappu2015,Marmol2015}. These measured EWs are higher than predicted by models of galaxy formation \citep[e.g.][]{Wilkins2013b} and the origin of these ubiquitous high-EW lines is still unclear and proves an ongoing challenge in our current understanding of the physical properties of high-redshift galaxies. 

While the interpretation of the [\ion{O}{3}] line strength is
complicated by the dependence on, for example, the gas density in the
\ion{H}{2} regions \citep[e.g.][]{Kewley2013,Shirazi2014}, the
H$\alpha$ line strength is known to be stable against variations in
density or temperature and therefore should be a stable tracer of the
star formation \citep{Kennicutt1998}. \citet{Shim2011} consider the
derived H$\alpha$ fluxes from a spectroscopic sample of sources in the
range $z=3.8$-5.0, where H$\alpha$ falls into the 3.6$\,\micron$
\textit{Spitzer}/IRAC filter, and find that the inferred H$\alpha$
flux in their spectroscopic sample is particularly elevated relative
to the inferred SFR from the UV continuum.  \citet{Shim2011} argue that one probable
explanation for the high ratio of H$\alpha$ flux to $UV$ flux they
observe could be due to a preference for young ages amongst their
selected sources, as essentially all sources with spectroscopic
redshifts that they consider show Ly$\alpha$ emission.  While this
speculation by \citet{Shim2011} is reasonable, the actual impact of considering 
only those sources showing Ly$\alpha$ emission is unclear; 
it requires testing based on a much larger and unbiased
sample of $z\sim4$ galaxies and one also benefitting from even deeper
photometric observations.

\begin{figure*}[]
\centering
\includegraphics[width=0.9\textwidth,trim=15mm 175mm 5mm 35mm] {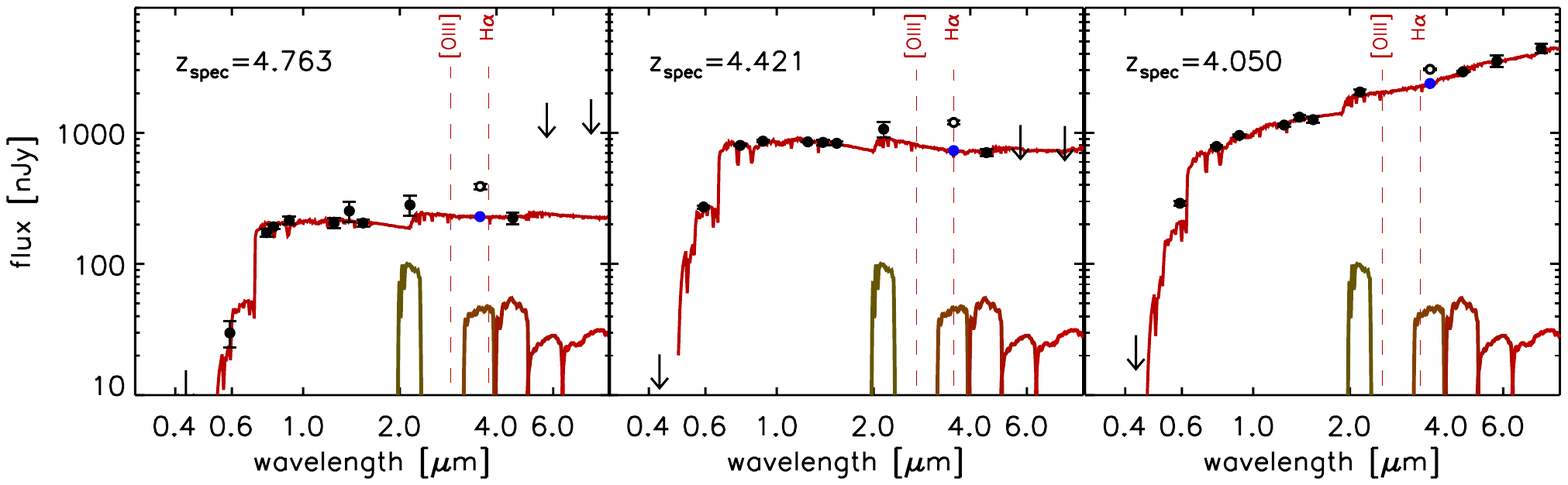} 
\caption{ Three examples of stellar population fits to the broadband observations of galaxies from our sample. Flux densities and upper limits (2$\sigma$) of the \textit{HST}, ground-based and \textit{Spitzer}/IRAC photometry are indicated with black points and arrows, while the best-fit stellar population (\S \ref{sec:SED}) is drawn in red. Filter transmission curves of the rest-frame optical bands are drawn to show the coverage of the SED. The IRAC 3.6$\,\mu$m band flux  is contaminated by  H$\alpha$, [\ion{N}{2}] and [\ion{S}{2}]  and is not included in the SED fitting (open points). The 4.5$\,\mu$m band is largely free of line contamination and therefore provides the most important constraint on the stellar continuum at rest-frame visible wavelengths. The offset between the predicted 3.6$\,\mu$m continuum flux from the SED (indicated by the blue points) and the observed 3.6$\,\mu$m flux, provides a good estimate of the total H$\alpha$+[\ion{N}{2}]+[\ion{S}{2}] line flux. 
 } 
\label{fig:examples}
\end{figure*}

In this paper we revisit the use of H$\alpha$ as a SFR indicator in
the redshift range $z=3.8$-5.0, considering both expanded
spectroscopic and photometric-redshift selections.  In doing so, we
leverage even deeper \textit{Spitzer}/IRAC coverage from the S-CANDELS
survey \citep{Ashby2015} and deep $K$-band data
\citep{Kajisawa2006,Hathi2012,Fontana2014}. This approach allows us to
make a state of the art assessment on the origin of high-EW H$\alpha$
emission in typical high redshift sources.

We search for correlations of the H$\alpha$ EW with a large number of
observational and physical properties, and we look for possible biases
in the results of spectroscopic samples relative to
photometric-redshift-selected samples and vice versa.  We use the
H$\alpha$ fluxes to derive to estimate specific star formation rates
from galaxies and compare these to sSFRs derived from the
$UV$-continuum fluxes in an effort to constrain the burstiness of the
star formation history.  Finally, we will discuss the implications of
our results for the so-called main sequence of star-forming galaxies
and the $z\sim4$-8 SFR functions.

This paper is organized as follows. In \S \ref{sec:Observations} we describe the observations we use and how we define our spectroscopic and photometric-redshift-selected samples, while we derive the observational and physical properties of our samples in \S \ref{sec:prop}. In \S \ref{sec:sfrHa} we derive H$\alpha$-based SFRs, which we compare with UV-based SFRs and we discuss the potential origin of the discrepancy we find between the different probes. In \S \ref{sec:mainseq} we establish the main sequence of star-forming galaxies from our H$\alpha$ measurements, while in \S \ref{sec:SFRfunc} we translate our findings into SFR functions. Finally, we summarize our results in \S \ref{sec:summary}.

Throughout this paper we adopt a Salpeter initial mass function (IMF) with limits 0.1-100$\,M_\odot$ \citep{Salpeter}. For ease of comparison with previous studies we take $H_0=70\,\rm km\,s^{-1}\,Mpc^{-1}$, $\Omega_{\rm{m}}=0.3$, and $\Omega_\Lambda=0.7$. Magnitudes are quoted in the AB system \citep{OkeGun}.

\begin{figure*}
\centering
\includegraphics[width=0.9\textwidth,trim=10mm 175mm 10mm 35mm] {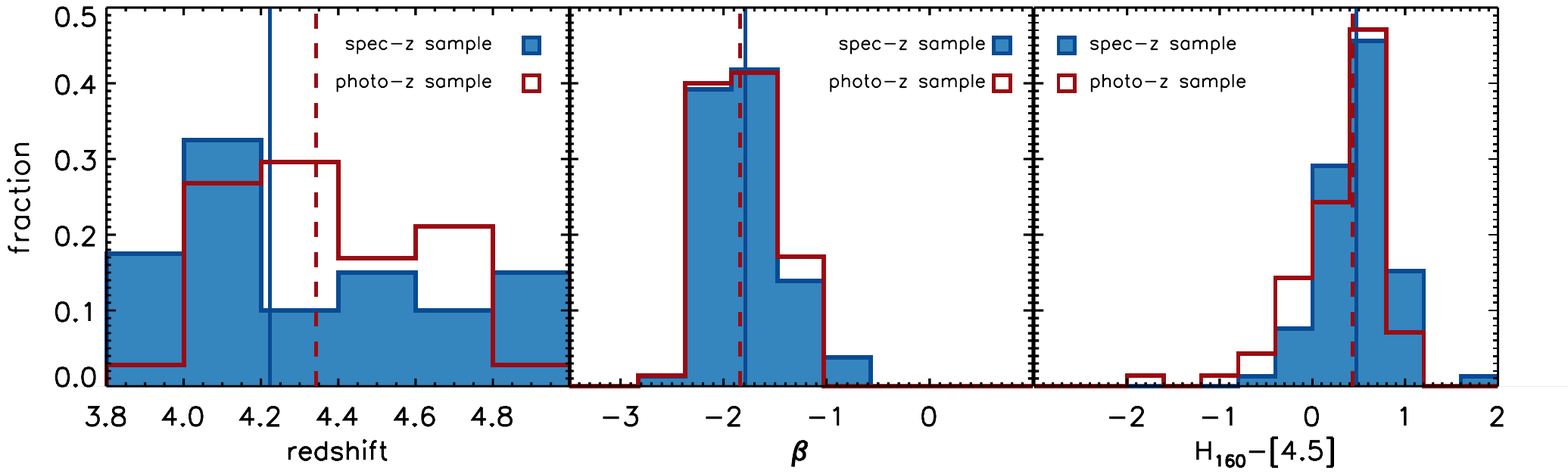} 
\caption{Comparison of the observational properties of $i_{775}$-band limited photometric-redshift-selected subsample (filled blue histograms) and spectroscopic-redshift-selected (red histograms) sample. The solid blue lines and dashed red lines indicate the median values for the spectroscopic and photometric-redshift-selected samples respectively. The median $i_{775}$-band magnitudes of the two samples are identical by construction, due to our $i_{775}\lesssim 25.4$ luminosity cut on the photometric-redshift-selected sample (see \S \ref{sec:phot_sample}).  \textit{Left panel:} The redshift distribution of the two samples. We find a median $<z_{\rm spec}>\sim4.25$ and a median $<z_{\rm phot}>\sim4.38$. \textit{Middle panel:} The UV-continuum slope, $\beta$, defined as $f_{\lambda}\propto \lambda^\beta$.  The median UV-continuum slope of the spectroscopic sample is slightly bluer than the median slope for the photometric-redshift-selected sample. \textit{Right panel:} The $H_{160}-[4.5]$ color, where the $H_{160}$-band probes the rest-frame UV-continuum and the 4.5$\,\micron$ band gives a measure of the rest-frame optical continuum flux. The medians of the two samples are within errors. Overall, the colors of the SEDs of the spectroscopic sample show very little bias when compared to the photometric-redshift-selected sample with the same median $i_{775}$-band luminosity, even though the majority of our spectroscopic redshifts are obtained from Ly$\alpha$ in emission.     }
\label{fig:spec_vs_phot}
\end{figure*}

\section{Data and Samples}
\label{sec:Observations}

\subsection{Spectroscopic Redshift Sample}
\label{sec:spec_sample}
For our main sample of $z\sim4$ galaxies we take advantage of the spectroscopic redshift information collected over the GOODS-N and GOODS-S by the public samples of \citet{Shim2011,Stark2013,Balestra2010,Vanzella2005,Vanzella2006,Vanzella2008} and \citet{Vanzella2009}. These authors have collected galaxy samples from spectroscopic follow-up of $B$- and $V$-drop selected galaxy candidates, typically using $i_{775}$ or $I_{C}$ as the detection band. Redshifts for these galaxies are mainly derived from the position of the Ly$\alpha$ emission line, although redshifts for a few bright galaxy candidates are derived from their UV absorption lines or continuum breaks.

We select sources with secure spectroscopic redshifts between $z=3.8$ and $z=5.0$; the redshift range where the H$\alpha$ line contributes to the flux in the 3.6$\,\mu$m band, while the 4.5$\,\mu$m band is free of contamination from strong nebular lines \citep[see][]{Shim2011,Stark2013}. Within this redshift range the $K$-band is largely free of strong emission lines such as [\ion{O}{3}], H$\alpha$ and H$\beta$, though the [\ion{O}{2}]$\lambda3727${\AA} emission line could result in a boost to the $K$-band flux ($\sim0.1-0.2$ mag) for galaxies between $z=4.35$ and $z=5.0$ (affecting 43\% of our sample).

We obtain photometry for the sources in our sample by matching the spectroscopic $z=3.8$-5.0 sample with the public 3D-\textit{HST}/CANDELS catalogs presented by \citet{Skelton2014}. We utilize their measured photometry in all \textit{HST} bands ($B_{435},V_{606},i_{775},I_{814},z_{850},J_{125},JH_{140}$ and $H_{160}$). The median 5$\sigma$ limiting magnitude in the bands, measured in a 0\farcs7 diameter aperture, ranges from 25.6 to 27.4. In short, \citet{Skelton2014} obtain their photometry by running Source Extractor \citep{Bertin1996} in dual image mode on all bands, while matching their point spread function (PSF) to the $H_{160}$-band PSF. A  combination of the $J_{125},JH_{140}$ and $H_{160}$ images is used as the detection image (weighted by the square root of the inverse variance) and total fluxes are measured in Kron apertures.

Furthermore, we include the photometry in the \textit{Spitzer}/IRAC bands at 5.8 and 8.0$\,\mu$m from the \citet{Skelton2014} photometric catalogs, who make use of the GOODS Spitzer 3rd data release. We obtain the deepest possible photometry for the 3.6$\,\mu$m and 4.5$\,\mu$m bands by leveraging the imaging from the  \textit{Spitzer} Extended Deep Survey \citep[SEDS:][]{Ashby2013} and the  \textit{Spitzer} Very Deep Survey Exploration Science Project \citep[S-CANDELS:][]{Ashby2015}  which covers the GOODS-N and GOODS-S fields with up to 50 hours exposure times (26.8 mag at 5$\sigma$ in a 2\farcs0 diameter aperture in the 3.6$\,\micron$ band). Similar to the procedure used for the public 3D-\textit{HST}/CANDELS catalogs described in \citet{Skelton2014} we obtain photometry with the Multi-resolution Object PHotometry oN Galaxy Observations (MOPHONGO) code described in \citet{Labbe2006,Labbe2010a,Labbe2010b,Labbe2015}, which provides an automated cleaning procedure for deblending the sources of interest and their neighboring sources. In short MOPHONGO creates model fluxes for all sources in a $\sim11\arcsec$ radius by PSF-matching all detected galaxies in the $H_{160}$ band to the IRAC image PSF and simultaneous fitting the normalizations of the modeled galaxies to match the observed IRAC image. Cleaned images are created by subtracting the model fluxes of all neighboring sources from the observed image. We measure the flux in the $3.6$ and $4.5\,\micron$ bands from the cleaned images in 2\farcs0 diameter apertures and we apply a $\sim2.2-2.4\times$ aperture correction based on the ratio of the flux enclosed in the photometric aperture in the \textit{HST} image (before convolution) to the IRAC model (after convolution) of the source of interest.

In order to obtain good constraints on the rest-frame optical stellar light, both shortwards and longwards of the H$\alpha$ emission line, we require good signal-to-noise $K$-band photometry.  For the GOODS-N field we  therefore combine CFHT/WIRCam $K_s$-band imaging \citep{Hathi2012} and Subaru/MOIRCS $K_s$-band imaging \citep{Kajisawa2006}. For the GOODS-S field we combine deep FOURSTAR $K_s$-band imaging from the Z-FOURGE survey \citep{Spitler2012}, VLT/ISAAC $K_s$-band imaging from the FIREWORKS survey \citep{Retzlaff2010} and VLT/HAWK-I K$_s$ band imaging from the HUGS survey \citep{Fontana2014}. We use MOPHONGO to perform an identical deblending procedure as described in the previous paragraph and we perform photometry on the cleaned images in 1\farcs0 diameter apertures. The median 5$\sigma$ limiting depths are 24.8 and 25.2 mags (in a 1\farcs0 diameter aperture) in GOODS-N and GOODS-S respectively; 44\% of our spectroscopic sample is detected at $>5\sigma$ in the  $K_s$-band. 

Our resulting catalog of $z_{\rm spec}=3.8$-5.0 galaxies consists of 37 sources in GOODS-N and 53 sources in GOODS-S, with high-quality constraints on the spectral energy distributions (SEDs) of the galaxies. The SEDs of three typical galaxies are presented in Figure \ref{fig:examples}.

\subsection{Photometric Redshift Sample}
\label{sec:phot_sample}

We complement our spectroscopic redshift sample with a high-confidence
photometric sample to add valuable statistics.  Use of a
photometric-redshift-selected sample is also valuable to evaluate
potential biases in the spectroscopic sample that may arise due to
these samples being predominantly composed of galaxies that show
Ly$\alpha$ in emission.  For this photometric sample we utilize the
public photometric redshift catalog over the GOODS-S field from the
3D-\textit{HST}/CANDELS data release \citep{Skelton2014}, generated
using the EAZY software \citep{Brammer2008}. We require selected
sources to have a photometric redshift within the redshift range
$z=3.8$-5.0 with at least 99\% probability.  Of the sample of sources
 that satisfy the 99\% probability criterion,  84\% has a best-fit photometric
  redshift template with a reduced $\chi^2$ value of less then 3 (95\% has a $\chi_{\rm red}^2<5$). 
  Of the sources in the 
spectroscopic sample, 71\% satisfy this criterion, with the remaining 29\% 
having a spectroscopic redshift at the edges of the $z=3.8$-5.0 redshift range used for selection.

We assemble the photometry
for our photometric sample in an identical way to the spectroscopic
sample as described in \S \ref{sec:spec_sample}. We apply a luminosity
cut for our main sample below $H_{160} < 26.5$ in order to ensure
high-S/N IRAC photometry for our entire sample. This results in a
photometric catalog containing 320 sources. Furthermore, we include an
$i_{775}$ limited subsample that we use to investigate the bias of our
spectroscopic sample in the derived galaxy properties, with respect to
our photometrically selected sources. For this subsample we use
$i_{775} < 25.4$ (80 sources) in order to match the median $i$-band
luminosity of our photometric as closely as possible to the median of
the spectroscopic sample ($i_{775}=25.1$). In the main photometric-redshift-selected
sample, 64 sources are included that are also part of the spectroscopic sample. 
For the remainder of the paper, we quote numbers for the separate selections 
with these spectroscopically-confirmed sources included. However, when quoting 
measured quantities for the combined sample, we only count a given source once (even if in both samples).

We compare the observational properties of our $i_{775}$-band limited
photometric-redshift-selected subsample with the spectroscopic sample
in Figure \ref{fig:spec_vs_phot}. The UV-continuum colors are
parametrized using the UV-continuum slope $\beta$, with
$f_{\lambda}\propto \lambda^\beta$. The $\beta$-slope is approximated
by a log-linear fit to the $z_{850}, J_{125},JH_{140}$ and $H_{160}$
fluxes \citep{Bouwens2012,Castellano2012}. The difference in the median 
UV-continuum color and $H_{160}-[4.5]$ color between our  $i_{775}$-band 
limited photometric subsample and our spectroscopic sample are consistent 
within the bootstrapped uncertainties. We therefore conclude that the spectroscopic sample
targeting mainly Ly$\alpha$ emitters has no obvious bias
 with respect to a photometric-redshift-selected sample
given similar $i_{775}$-band luminosities \citep[see also][]{Schenker2013}.
  
In \S\ref{sec:SED} we remove those sources with bad SED fitting, which results in a final sample size of 80 sources in the spectroscopic and 302 sources in the photometric catalog. We tabulate the properties of our final spectroscopic and photometric-redshift selected samples in Table \ref{tab:sample}.

\begin{figure*}
\centering
\includegraphics[width=0.85\textwidth,trim=10mm 175mm 10mm 40mm] {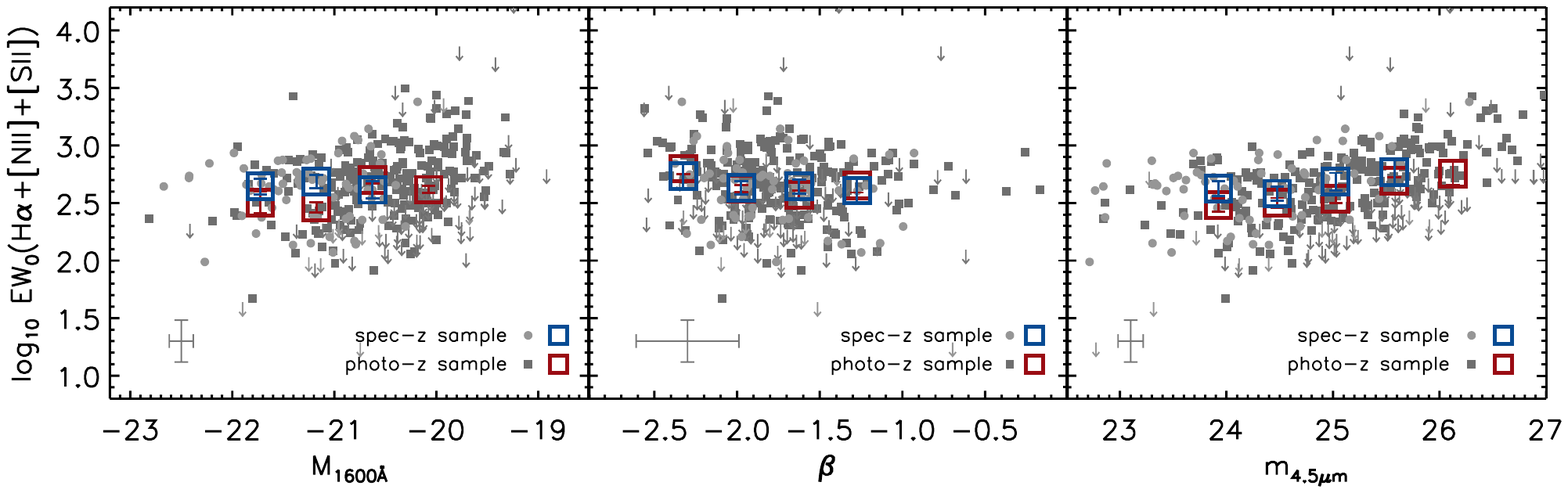} 
\includegraphics[width=0.85\textwidth,trim=0mm 177mm 20mm 37mm] {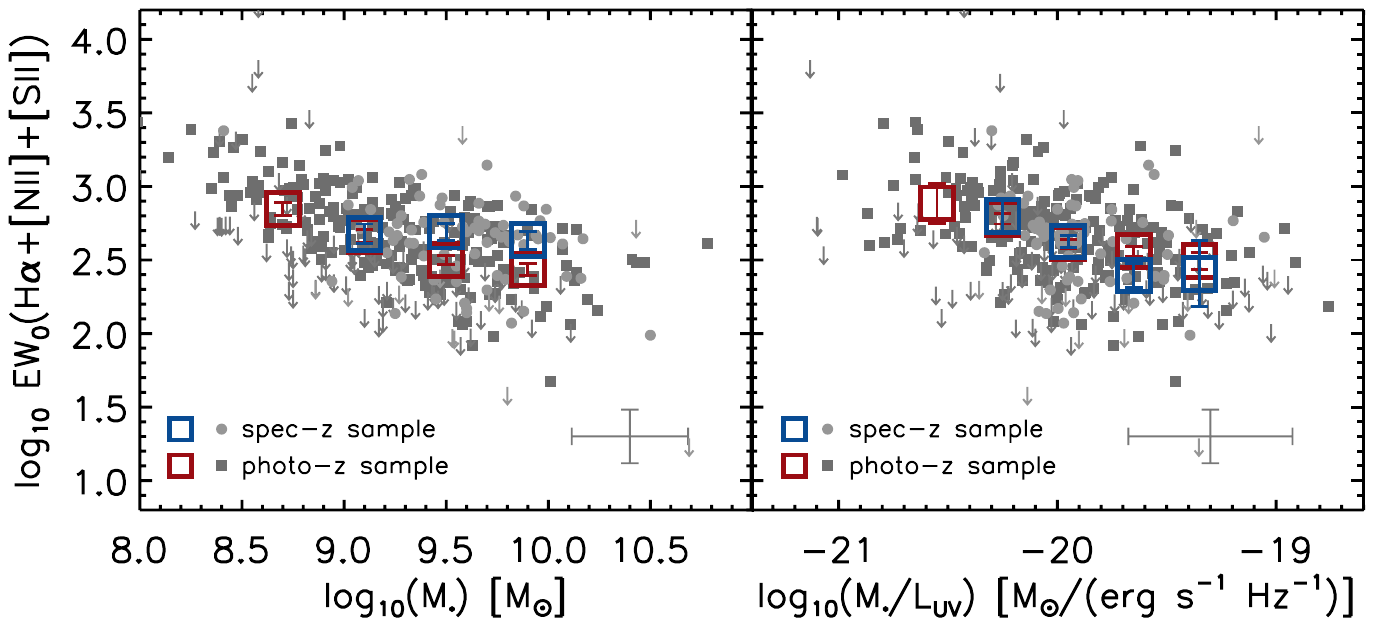} 
\caption{The rest-frame H$\alpha$+[\ion{N}{2}]+[\ion{S}{2}] EWs of our spectroscopic (light grey points) and our main photometric-redshift-selected sample (dark grey squares), measured from the offsets in the observed 3.6$\,\micron$  flux with respect to the predicted stellar continuum (see \S \ref{sec:Ha}). Errorbars for individual points are not shown to improve  the clarity of the figures; instead representative errorbars are shown at the bottom of each panel. The left, middle and right panel show the dependence of EW$_0$(H$\alpha$+[\ion{N}{2}]+[\ion{S}{2}]) as a function of the observed UV luminosity (measured at 1600{\AA} from the best fit stellar template), UV-continuum slope $\beta$ and  observed $[4.5]$ magnitude respectively.
The blue and red squares indicate the median EWs (error bars represent the uncertainty in the median).  We explore the dependence  of the median emission line EW on the UV luminosity, UV-continuum slope $\beta$, observed [4.5] magnitude, stellar mass and mass-to-light ratio. The emission line EW is clearly correlated with $M_\ast/L_{\rm UV}$ in both the spectroscopic and photometric samples, indicating that the EWs might be mainly driven by the star-formation activity in the galaxy.  
} 
\label{fig:EWs}
\end{figure*}

\begin{table*}
\scalebox{0.97}{
\begin{threeparttable}
\centering
\caption{Median properties of the spectroscopic and photometric-redshift-selected samples}
\begin{tabular}{lccccccccc} 
\hline \hline 
 & N  & $z$  & $M_{\rm UV}$& $M_\ast$ & $\beta$ & EW$_{\rm 0}^a${\scriptsize (H$\alpha$)} & EW$_{\rm 0}^a${\scriptsize (H$\alpha$+[NII]+[SII])} & sSFR$_{\rm UV}^{a,b}$ & sSFR$_{\rm H\alpha}^{a,b}$\\
  &   &  & & [$M_\odot$] & &  [\AA] &  [\AA]  & [Gyr$^{-1}$] & [Gyr$^{-1}$] \\
\hline 
spec-z                                                      & 80 & 4.3 & $-21.1$ & $4.0\cdot 10^9$ & -1.79 & $361\pm19$ & $429\pm23$ & $13.3\pm0.6$ & $16.7\pm3.4$  \\
photo-z (all)                                           & 302 & 4.4 & $-20.3$ & $1.5\cdot 10^9$ & -1.79 & $335\pm22$ & $399\pm27$ & $15.1\pm1.1$ & $ 17.6\pm2.0$ \\
photo-z ($i_{775} < 25.4$)                    & 71 & 4.4 & $-21.1$ & $3.2\cdot 10^9$ & -1.84 & $269\pm80$ & $320\pm96$ &  $15.1\pm1.8$  & $14.5\pm4.6$ \\
photo-z (${\rm log_{10}} M_\ast>9.5$) & 88 & 4.4 & $-20.7$ & $5.5\cdot 10^9$ & -1.62 & $220\pm38$ & $262\pm46$ & $5.7\pm1.0$ & $5.8\pm1.1$ \\
\hline 
\end{tabular} 
\label{tab:sample}
$^a$ Measured median values and uncertainties obtained from bootstrapping.

$^b$ Corrected for dust using the UV slope $\beta$ and the \citet{Meurer1999} calibration; H$\alpha$ is corrected for dust assuming $A_{V,\rm stars}= A_{V, \rm gas}$ and using the \citet{Calzetti2000} dust curve. 
\end{threeparttable}
}
\end{table*}

\begin{table*}
\scalebox{1.0}{
\begin{threeparttable}
\centering
\caption{Dependence of EW$_0$(H$\alpha$+[\ion{N}{2}]+[\ion{S}{2}]) on observational properties$^a$}
\begin{tabular}{lcc}
\hline\hline 
 & spec-z & photo-z \\ 
\hline 
$d\,{\rm log_{10} EW_{0,H\alpha+[NII]+[SII]}}/d\, M_{\rm UV}$                 & 0.08$^{+0.13}_{-0.14}$ & 0.08$^{+0.04}_{-0.08}$ \\
$d\,{\rm log_{10} EW_{0,H\alpha+[NII]+[SII]}}/d\, \beta$                          &  0.08$^{+0.20}_{-0.20}$ &  $-$0.17$^{+0.06}_{-0.07}$ \\ 
$d\,{\rm log_{10} EW_{0,H\alpha+[NII]+[SII]}}/d\, m_{\rm 4.5\mu m}$              &  0.01$^{+0.09}_{-0.08}$ &  0.14$^{+0.05}_{-0.04}$\\
$d\,{\rm log_{10} EW_{0,H\alpha+[NII]+[SII]}}/d\,{\rm log_{10}}M_{\ast}$                      &  0.12$^{+0.17}_{-0.23}$ & $-$0.35$^{+0.08}_{-0.07}$ \\
$d\,{\rm log_{10} EW_{0,H\alpha+[NII]+[SII]}}/d\,{\rm log_{10}}(M_{\ast}/L_{\rm UV})$  &  $-$0.09$^{+0.25}_{-0.24}$ &  $-$0.36$^{+0.06}_{-0.07}$ \\
\hline
\end{tabular} 
\label{tab:EWs}
$^a$ Measured linear slopes from the median binned data. Uncertainties are obtained from bootstrapping the binned data and re-fitting a linear slope to each bootstrapped set of medians.
\end{threeparttable}
}
\end{table*}

\section{Derived properties of $\lowercase{z}=3.8$-5.0 galaxies}
\label{sec:prop}
\subsection{SED fitting}
\label{sec:SED}
We determine stellar masses and other stellar population parameters by fitting stellar population synthesis templates to the observed photometry using FAST \citep{Kriek2009}. We do not include emission lines in our galaxy templates; instead we consider only stellar continuum in our models while we exclude the 3.6$\,\micron$ band, where H$\alpha$,  [\ion{N}{2}] and [\ion{S}{2}] boosts the observed flux, from our fitting procedure. We consider constant star formation histories with ages between 10 Myr and the age of the Universe at $z=3.8$. Furthermore, we assume a \citet{Calzetti2000} dust law, with $A_V$ in the range $0-2$. Finally, we allow the metallicities to range between $0.2-1.0Z_\odot$ in the fits.  

We fix the redshifts in the SED fitting either to the spectroscopic redshift of the galaxies or to the photometric redshift value used to select galaxies in the photometric redshift sample (\S \ref{sec:phot_sample}). Note that when we let the redshift of the galaxies in the photometric sample float, the estimated median EW$_0$(H$\alpha$+[\ion{N}{2}]+[\ion{S}{2}]) changes by only $+0.01$ dex.
For a small number of galaxies we find a bad fit to the photometry and we therefore remove 9 sources from the spectroscopic sample and 18 sources  from the photometric redshift sample when the reduced $\chi^2$ is greater than four.

Our estimated median stellar mass is $4.4\cdot 10^9 \,M_\odot$ for our main spectroscopic sample and $1.6\cdot10^9 \,M_\odot$ for our photometric sample, reflecting the fact that our main photometric sample extends to lower luminosities.

\begin{figure}
\centering
\includegraphics[width=1.0\columnwidth,trim=110mm 125mm 25mm 35mm] {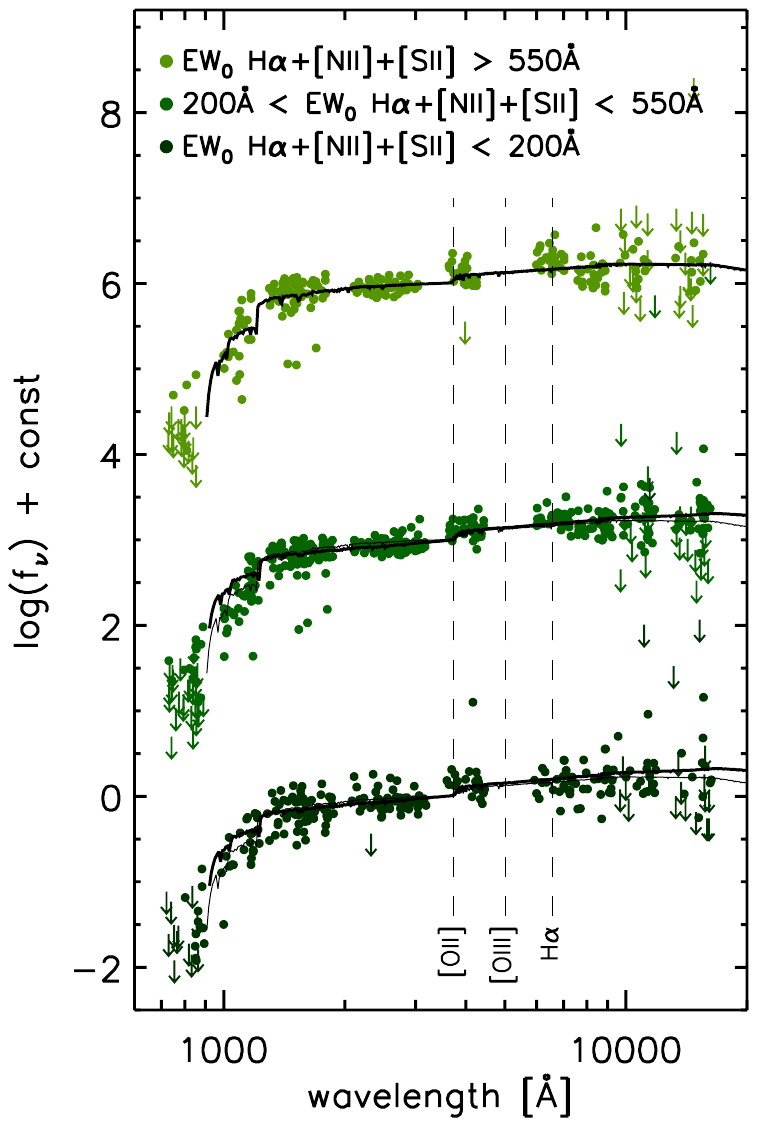} 
\caption{Composite spectral energy distributions of the sources in the spectroscopic sample, shown as a function of the rest-frame wavelengths (green points).  The sources presented in the top, middle, and bottom SED have estimated EW$_0$(H$\alpha$+[\ion{N}{2}]+[\ion{S}{2}]) that are greater than 550{\AA}, between 200{\AA} and 550{\AA}, and less than 200{\AA}, respectively. The SEDs are offset in the y-axis for clarity. All points are normalized by the log mean of the $z_{850}, H_{160}$ and $4.5\,\micron$ fluxes.  The dashed black lines indicate the position of the [\ion{O}{2}], [\ion{O}{3}] and H$\alpha$ nebular lines. The thick black curves indicate a stellar continuum fit to the composite SEDs.  The continuum SED for the highest EW sources is indicated with thin black curves next to the middle and bottom SEDs for reference. From the composite SEDs, high EW sources are slightly bluer and lower mass, consistent with the results in \S \ref{sec:Ha}.
 }
\label{fig:composite}
\end{figure}

\subsection{H$\alpha$+[\ion{N}{2}]+[\ion{S}{2}] equivalent widths and line strengths }
\label{sec:Ha}
We infer the total emission line flux in our sources from the
3.6$\,\micron$ band by subtracting the predicted continuum fluxes of
the best fit stellar templates (see Figure \ref{fig:examples}) from
the observed 3.6$\,\micron$ fluxes.  A correction is made for the
width of the 3.6$\,\micron$ filter using the spectral response curve
of this
filter\footnote{http://irsa.ipac.caltech.edu/data/SPITZER/docs/
  irac/calibrationfiles/spectralresponse/} \citep[see
  also][]{Shim2011,Stark2013}. We estimate the uncertainty on the
predicted continuum flux to be equal to the uncertainty on the
4.5$\,\micron$ band flux, and therefore the uncertainty on the flux offset
 in the 3.6$\micron$ band is equal to the uncertainty in the [3.6]$-$[4.5] color.
 If the uncertainty on the line flux is larger than the offset between the
observed 3.6$\,\micron$ flux and the predicted continuum, we place an
upper limit on the line flux.

Since the total emission line flux is dominated by the contribution
from H$\alpha$, [\ion{N}{2}] and [\ion{S}{2}]
\citep[e.g.][]{Anders2003} we directly obtain rest-frame equivalent
widths (EW$_0$) for these lines based on the inferred total emission
line flux and the predicted continuum of the best-fit stellar
template, after correcting the observed EW by a factor $(1+z)$. Here,
we use the FAST redshift estimates for the photometric sample. We find
EW$_0$(H$\alpha$+[\ion{N}{2}]+[\ion{S}{2}]) $\sim399$-429{\AA} in the
median source of our samples. We did not correct the
continuum emission or line emission for dust attenuation.

Our estimate of the equivalent width is in good agreement with
\citet{Stark2013}, who measure $<\log_{10}(\rm
EW_{3.6\,\micron})>\sim2.57-2.73$ in the rest-frame.  The present
result is $\sim$20\% higher than recent results by \citet{Marmol2015},
but this may be due to the fact the median stellar mass for our sample
is $\sim$0.9 dex lower than the sample considered by
\citet{Marmol2015} and a possible correlation of the H$\alpha$ EW with
stellar mass \citep[EW$_{H\alpha}\propto
  M_{*}^{-0.25}$:][]{Fumagalli2012,Sobral2014}.\footnote{We remark,
  however, that the apparent correlation of the H$\alpha$ EW with
  stellar mass could be significantly impacted by source selection and
  the fact that the lowest sSFR, lowest-mass sources simply could not
  be selected and included in current samples.  See
  Figure~\ref{fig:mainseq}.}  Furthermore, our estimate is
lower than the $z\sim5$ estimate of the equivalent width by
\citet{Rasappu2015}, who derive $\sim665$ {\AA} from the median
$[3.6]-[4.5]$ color. The difference between the $z\sim4.3$ to
$z\sim5.2$ equivalent width estimates is consistent with an evolution
of $EW_0\propto(1+z)^{1.8}$ within $<2\sigma$ \citep{Fumagalli2012,Sobral2014}.
\citet{Shim2011} derive a much higher
 EW$_0$(H$\alpha$+[\ion{N}{2}]+[\ion{S}{2}])$\sim600${\AA} over the same
 redshift range, using a similar method of deriving H$\alpha$ fluxes. We have 52 sources in 
 common with the \citet{Shim2011} sample, 
 but using our method we find a median EW$_0$(H$\alpha$+[\ion{N}{2}]+[\ion{S}{2}])$=416${\AA}
  for these sources. We find that our lower EW measurements for the same sources are possibly due to a systematically lower
3.6$\micron$ flux ($\Delta([3.6]-[4.5])=0.21\pm0.08$ mag) than used by \citet{Shim2011}.
  \citet{Marmol2015} also find a 0.2 mag discrepancy with the  3.6$\micron$ photometry used by \citet{Shim2011}.

We show the distribution of the resulting EWs as a function of the UV-luminosity (measured at 1600{\AA} from the best fit stellar template), UV-continuum slope $\beta$, observed 4.5 $\mu m$-band  magnitude in the top panels of Figure \ref{fig:EWs}.  We use a linear fit to the median bins in Figure \ref{fig:EWs}, where we bootstrap every bin 1000 times and re-fit a linear relation to obtain realistic errors on the linear slope. We list the slopes and bootstrapped uncertainties in Table \ref{tab:EWs}. 
With this method, we find that that EW$_0$(H$\alpha$+[\ion{N}{2}]+[\ion{S}{2}]) in our spectroscopic sample is consistent (at $\lesssim2\sigma$) with no correlation for all distributions in Figure \ref{fig:EWs}. In the photometric-redshift-selected sample, the typical derived EW$_0$(H$\alpha$+[\ion{N}{2}]+[\ion{S}{2}])  seems to be weakly dependent $\beta$ and observed 4.5 $\mu m$-band  magnitude (at the $\sim2$-3$\sigma$ level). 

 To gain further insight into the possible physical origin for these
 high EW lines, we use the public catalogs with structural parameters
 presented by \citet{vanderwel2012,vanderwel2014} to identify
 potential correlations with our H$\alpha$ measurements. However, we
 find no dependence of EW$_0$(H$\alpha$+[\ion{N}{2}]+[\ion{S}{2}]) on
 half-light radius or S\'{e}rsic index. 
 
 Furthermore, we study the
 dependence of H$\alpha$ EW on the parameters from our stellar
 population modeling (\S \ref{sec:SED}) in the bottom panels of Figure \ref{fig:EWs}. 
 For our photometric-redshift-selected sample we find a dependence
 of EW$_0$(H$\alpha$+[\ion{N}{2}]+[\ion{S}{2}]) on 
 stellar mass and $M_\ast/L_{\rm UV}$ (at $\gtrsim5\sigma$), which is expected if the H$\alpha$ is predominantly
 determined by the star-formation activity in the galaxy. 

 Therefore we will explore the use of our inferred H$\alpha$
 measurements as a star-formation rate indicator in the next section.
 To do this, we derive an estimate of the H$\alpha$ line flux by
 adopting a fixed ratio between H$\alpha$, [\ion{N}{2}] and
 [\ion{S}{2}] as tabulated in \citet{Anders2003} for sub-solar
 ($0.2Z_\odot$) metallicity, i.e.  $L_{\rm H\alpha}=0.84\times
 L_{3.6\,\micron}$, where  $L_{3.6\,\micron}$ is the total 
 luminosity derive from the offset in the 3.6 $\mu m$ band with
  respect to the estimated continuum from the SED. 
  This is consistent with the findings of
 \citet{Sanders2015}, who observe a ratio of \ion{N}{2}/H$\alpha$ of
 0.05--0.09 in $z\sim2.3$ galaxies with stellar masses in the range
 $\log(M_\ast/M_\odot) = 9.15-9.94$.  The resulting H$\alpha$ EW is
 361{\AA} for our spectroscopic and 335{\AA} for our photometric
 redshift selected sample.

\begin{figure*}
\centering
\includegraphics[width=1.\textwidth,trim=17mm 175mm 33mm 35mm] {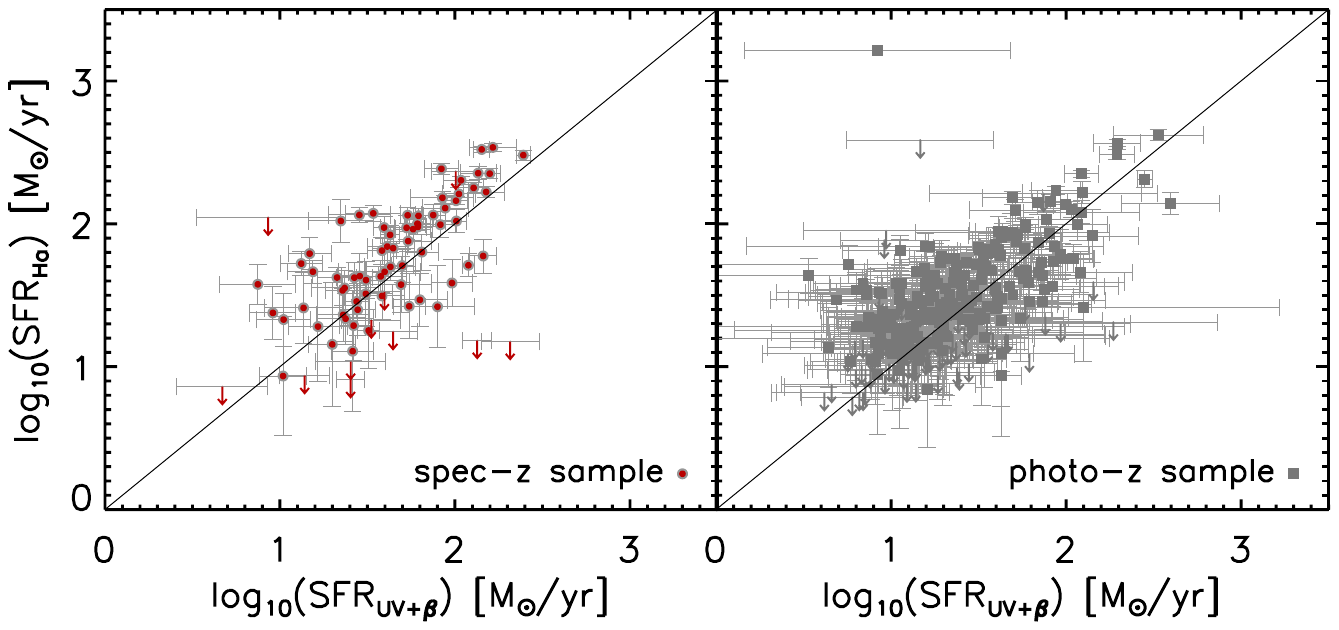} 
\caption{Star formation rates from the inferred H$\alpha$ luminosities versus those from the UV-luminosity corrected for dust using the UV slope $\beta$ and the \citet{Meurer1999} calibration (red points; red arrows indicate the 1$\sigma$ upper limits); H$\alpha$ is corrected for dust assuming $A_{V,\rm stars}= A_{V, \rm gas}$ and using the \citet{Calzetti2000} dust curve.  The left panel shows our spectroscopic sample, while the right panel shows our photometric sample. The median SFRs are offset from the one to one relation (black line) by $\sim$0.15 dex. } 
\label{fig:SFR_comp}
\end{figure*}

\begin{figure}
\centering
\includegraphics[width=0.85\columnwidth,trim=115mm 178mm 25mm 33mm] {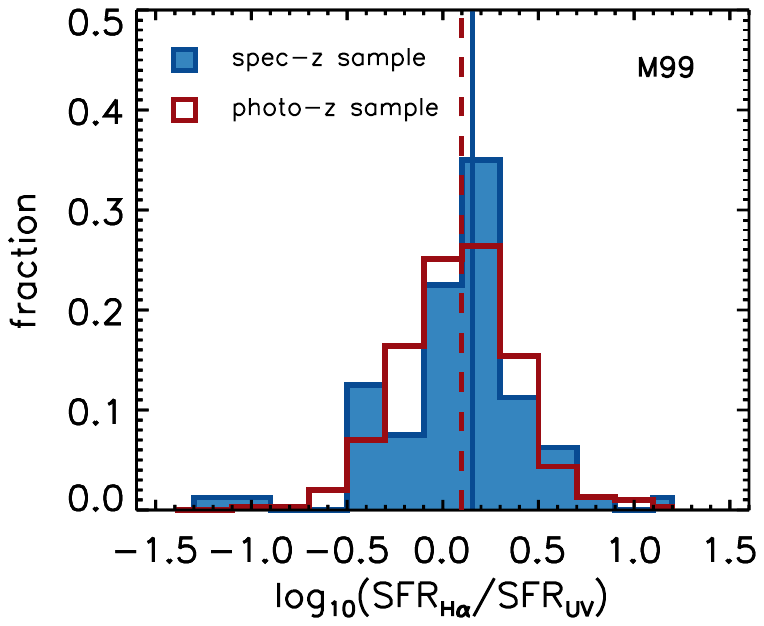} 
\caption{Histogram of star formation rates from the inferred H$\alpha$ luminosities versus those from the UV-luminosity corrected for dust using the UV slope $\beta$ and the \citet{Meurer1999} calibration for the spectroscopic sample (blue filled histogram) and photometric-redshift selected sample (red histogram). } 
\label{fig:hist_SFR_comp}
\end{figure}

\subsection{Composite SEDs}
\label{sec:composite}

Our spectroscopic sample is particularly well suited to construct composite SEDs of star-forming galaxies over the redshift range  $z=3.8$-5.0. Figure \ref{fig:composite} shows three composite SEDs, with galaxies divided in to different samples based their estimated EW$_0$(H$\alpha$+[\ion{N}{2}]+[\ion{S}{2}]). The flux measurements are normalized on the $z_{850}, H_{160}$ and $4.5\,\micron$ bands. By construction the offset in the flux measurements around the H$\alpha$ line increases from bottom to top. We find the highest EW$_0$(H$\alpha$+[\ion{N}{2}]+[\ion{S}{2}]) to have a slightly bluer and lower mass SED, consistent with the relations in Table \ref{tab:EWs} for the photometric sample.

\section{Inferred H$\alpha$ as star formation rate indicator}
\label{sec:sfrHa}

Our determination of the H$\alpha$ line flux in \S \ref{sec:Ha} provides us with the unique opportunity to explore the use of inferred H$\alpha$ fluxes as an independent star formation rate indicator in high-redshift galaxies. \citet{Shim2011} pioneered the use of inferred H$\alpha$ to measure SFRs of $z\sim4$ galaxies. However, the exceptionally deep S/N  Spitzer/IRAC data from the S-CANDELS data-set \citep{Ashby2015} covering our large spectroscopic and photometric-redshift-selected samples allows us to systematically assess the SFRs over the general $z\sim4$ galaxy population.   

\subsection{Star formation rate indicators}
\label{sec:SFRs}

In this section we will define two independent SFR indicators, based on the H$\alpha$ and UV properties of our samples, using calibrations of local  star-forming galaxies. The comparison of these two probes will allow us to investigate the different time scales of star-formation, since H$\alpha$ is sensitive to the star-formation history (SFH) over a $\sim$10 Myr timescale, while UV light provides a time-averaged SFR over a $\sim$100 Myr time window \citep[e.g.][]{Kennicutt1998}.  

To obtain UV-based SFRs we convert the UV-luminosity measured at 1600{\AA} from the best fit stellar template into a SFR using the \citet{Kennicutt1998} relation 
\begin{equation}
\label{Eq:Kennicutt_UV}
\rm SFR \,(M_\odot\,yr^{-1}) = 1.4 \times 10^{-28}\, \textit{L}_{UV} \,(erg\,s^{-1}\,Hz^{-1}).
\end{equation}
We estimate the dust attenuation in the UV from the calibration by \citet{Meurer1999} using local starbursting systems
\begin{equation}
\label{Eq:Meurer}
A_{1600} = 4.43 + 1.99 \cdot\beta,
\end{equation}
where we estimate the UV-continuum slope $\beta$ using a log-linear fit to the $z_{850}, J_{125},JH_{140}$ and $H_{160}$ fluxes.
Our estimated median dust corrected UV-based SFR ($\rm SFR_{UV+\beta}$) is equal to $\sim 43 \,M_\odot\,\rm yr^{-1}$ for our spectroscopic-redshift-selected sample and $\sim 20 \,M_\odot\,\rm yr^{-1}$ for our photometric-redshift-selected sample. We explicitly do not estimate our UV-based SFRs from the SED fitting procedure due to the degeneracy between age and dust that is particularly challenging to solve. However, we will discuss the impact of different calibrations of the dust law on the UV-based SFRs in \S \ref{sec:origin_dust}.

To obtain SFRs from the H$\alpha$ line luminosity measurements derived in \S \ref{sec:Ha} we use the \citet{Kennicutt1998} relation 
\begin{equation}
\label{Eq:Kennicutt_Ha}
\rm SFR \,(M_\odot\,yr^{-1}) = 7.9 \times 10^{-42}\,\textit{L}_{H\alpha} \,(erg\,s^{-1}).
\end{equation}
We estimate the dust attenuation from the \citet{Calzetti2000} dust law and the UV dust attenuation derived using Eq. \ref{Eq:Meurer}. Here, we assume $A_{V,\rm stars}= A_{V, \rm gas}$, which is expected to be a reasonable assumption for blue galaxies where both the stars and emission lines are in the birth clouds and which is found to be a reasonable approximation in $z\sim2$ galaxies \citep[e.g.][]{Erb2006,Reddy2010,Shivaei2015}. Local observations of star-bursting systems indicate $A_{V,\rm stars}= 0.44 \cdot A_{V, \rm gas}$  \citep[e.g.][]{Calzetti1997}.  However,  as we will show below our H$\alpha$ SFRs are already relatively high compared to our $UV$-continuum-based SFRs, a discrepancy that would significantly increase if we made the assumption that $A_{V,\rm stars}= 0.44 \cdot A_{V, \rm gas}$. Our estimated median SFR from H$\alpha$ ($\rm SFR_{H\alpha}$) after dust correction is equal to $\sim 50 \,M_\odot\,\rm yr^{-1}$ for our spectroscopic sample and $\sim 23 \,M_\odot\,\rm yr^{-1}$ for our photometric sample.

We compare $\rm SFR_{UV+\beta}$ and  $\rm SFR_{H\alpha}$ indicators in Figure \ref{fig:SFR_comp} and we find that the two indicators are strongly correlated in both the spectroscopic and photometric samples. This strong correlation suggests that our method of inferring H$\alpha$ line measurements from the broadband IRAC photometry can be used as tracer of star formation. However, we also find a systematic offset of $0.16^{+0.03}_{-0.04}$ dex for our spectroscopic and $0.10^{+0.03}_{-0.01}$ dex for our photometric-redshift selected sample (see also Fig. \ref{fig:hist_SFR_comp}) from the one to one relation (uncertainties obtained from bootstrapping). Had we assumed a differential dust attenuation between nebular light and star light, i.e. $A_{V,\rm stars}= 0.44 \cdot A_{V, \rm gas}$, these offsets would have increased to $\sim0.3$ dex.

We have investigated potential systematics in our method of obtaining H$\alpha$ flux measurements that could explain the offset between $\rm SFR_{UV+\beta}$ and  $\rm SFR_{H\alpha}$. In appendix A we present two tests performed at $z<3.8$ to check for systematics in the photometry and SED fitting used in this paper. First, we compared spectroscopically measured emission lines with line fluxes inferred from the broad-band photometry with the same method we use for our $z=3.8-5.0$ samples. We use H$\alpha$+[\ion{N}{2}] fluxes from galaxies in the 3D-HST grism survey at  $z=1.3-1.5$ \citep{Momcheva2016}. We find no significant offset in the median sources between spectroscopically measured line fluxes and the flux measurements inferred from the photometry.

As a second test we use sources from the GOODS-S spectroscopic catalogues described in \S\ref{sec:spec_sample} within the redshift range $z=3.0-3.8$, where both the 3.6$\micron$ and 4.5$\micron$ bands are uncontaminated by line flux from strong emission lines. We perform the exact same steps as describe for our $z=3.8-5.0$ spectroscopic galaxy sample in deriving the difference between the 3.6$\micron$ photometry and the predicted continuum flux from the SED fitting. While any major systematics in our photometry would result in a median offset between the two, we find only a minor negative offset ($\Delta[3.6]= -0.02\pm0.01$ dex)s, indicating that, if anything, the H$\alpha$ fluxes we derive are underestimated compared to their true value and therefore the discrepancy between $\rm SFR_{UV+\beta}$ and  $\rm SFR_{H\alpha}$ can only increase. 

Another systematic we need to consider is the influence of [\ion{O}{2}] on the K-band flux for the fraction of sources in the redshift range $z=4.35-5.0$. When we exclude the $K_s$-band photometry from the SED-fitting for galaxies from the spectroscopic sample in this redshift range, these sources have higher H$\alpha$+[\ion{N}{2}]+[\ion{S}{2}] EW by 0.05 dex. However, we find only a 0.002 dex difference in the median H$\alpha$+[\ion{N}{2}]+[\ion{S}{2}] EW of the total sample when specifically excluding the $K_s$-band for the $z=4.35-5.0$ sources. However, we note that again, the systematic influence of [\ion{O}{2}] can only increase the differences we see in between $\rm SFR_{UV+\beta}$ and  $\rm SFR_{H\alpha}$, not resolve the discrepancy.

The discrepancy between $\rm SFR_{UV+\beta}$ and  $\rm SFR_{H\alpha}$ was already noted by \citet{Shim2011}, who found a mean $\left<\rm SFR_{\rm H\alpha}/SFR_{\rm UV}\right>\sim6$, assuming no dust correction. This is significantly larger than the $\rm SFR_{\rm H\alpha}/SFR_{\rm UV}\sim2.1$ we find from our sample before dust correction. However the \citet{Shim2011} H$\alpha$ emitter sample is IRAC excess selected and could therefore be biased towards high inferred H$\alpha$ EWs. 

\section{Reconciling $H\alpha$ and $UV$-continuum-based SFRs}

In this section, we will discuss how the physical assumptions we make
regarding the dust law, the star-formation histories, and also
ionizing photon production efficiencies of $z\sim4$ star-forming
galaxies impact the SFRs we derive from H$\alpha$ and $UV$-continuum
emission.  As we demonstrated in the previous section, the use of
relatively standard assumptions (\citet{Calzetti2000} dust law and
$A_{V,\rm stars}= A_{V, \rm gas}$) results in a systematic $\sim-0.10-0.16$
dex offset between H$\alpha$-based SFRs and UV-based SFRs.

\begin{table*}
\begin{threeparttable}
\centering
\caption{Quantitative Considerations in Achieving Consistent SFR$_{\rm H\alpha}$ and SFR$_{\rm UV}$ measurements}
\begin{tabular}{lccc} 
\hline 
\hline
 & \multicolumn{3}{c}{Consistency of SFRs for dust model}\\ 
\hline
Assumed dust correction & $\log_{10}(\rm SFR_{\rm H\alpha}/SFR_{\rm UV})^a$ & $0.4 A_{\rm H\alpha}^{\dagger}$ & $0.4 A_{\rm UV}^{\dagger}$ \\
\hline
None$^b$ 																										& $0.35^{+0.04}_{-0.05}$&  0.00 & 0.00 \\
Meurer+99, $A_{V,\rm stars}= A_{V, \rm gas}\,^b$										& $0.16^{+0.02}_{-0.04}$& 0.12 & 0.35 \\
Meurer+99, $A_{V,\rm stars}= 0.44 \cdot A_{V, \rm gas}\,^b$ 					& $0.28^{+0.02}_{-0.01}$& 0.27 & 0.35\\
SMC dust correction, $A_{V,\rm stars}= A_{V, \rm gas}\,^c$ 						& $0.19^{+0.02}_{-0.02}$ & 0.04 & 0.19 \\
$A_{1600}= 1.99 (\beta+2.54)$, $A_{V,\rm stars}= A_{V, \rm gas}\,^c$ & $0.01^{+0.03}_{-0.04}$ & 0.19 & 0.57 \\
\\
\hline
\hline
\multicolumn{4}{c}{Other Physical Assumptions that Impact the Consistency of SFRs}\\ 
\hline 
Assumed Properties of Stars / SF history & $\Delta\log_{10}(\rm SFR_{\rm H\alpha}/SFR_{\rm UV})^g$ & $\Delta\log_{10}(\rm SFR_{\rm H\alpha})^g$ & $\Delta\log_{10}(\rm SFR_{\rm UV})^g$\\
\hline
Rising SFH (Reddy+2012)$^{d,f}$ & $+0.07$ & $+0.00$ & $-0.07$\\
Stellar rotation (Leitherer+2014)$^{e,f}$ & $+0.14$& $+0.24$ & $+0.10$\\
Stellar binaries (Eldridge\&Stanway 2014)$^{e,f}$ & $+0.31$& $+0.46$& $+0.15$\\

\hline
\end{tabular} 
\label{tab:summary}
$^{\dagger}$ Median estimated extinction (dex) in the $UV$-continuum (1600\AA) and H$\alpha$ line.\\
$^a$ Measured median values and bootstrapping uncertainties are based on the spectroscopic sample. The measured values are somewhat lower than those derived by \citet{Shim2011} who find $\log_{10}(\rm SFR_{\rm H\alpha}/SFR_{\rm UV})\sim0.78$ dex applying no dust correction, but in good agreement with the sSFRs found by \citet{Stark2013} and \citet{Marmol2015} who assume \citet{Meurer1999} dust corrections. \\
$^b$ See \S\ref{sec:SFRs}. \\
$^c$ See \S\ref{sec:origin_dust}.\\ 
$^d$ See \S\ref{sec:origin_rising}.\\
$^e$ See \S\ref{sec:origin_ionrad}.\\
$^f$ Determined for a stellar population with an age of 100 Myr. \\
$^g$ Value of the Inferred SFRs (and SFR ratios) using the \citet{Kennicutt1998} relations minus the actual SFRs.  If the value in the table is positive, SFRs (or SFR ratios) estimated from the observations (H$\alpha$ or UV light) using the \citet{Kennicutt1998} relations will be overestimated.  If the value in the table is negative, SFRs estimated using the \citet{Kennicutt1998} relations will be underestimated.\\
\end{threeparttable}
\end{table*}

\subsection{Dust law}
\label{sec:origin_dust}


In the previous section, we showed that UV and H$\alpha$ based SFR indicators are strongly correlated, though somewhat systemically offset ($\sim0.15$ dex) from a one-to-one relation, when using a \citet{Meurer1999} dust correction and assuming  the same level of dust extinction for nebular light as stellar light (i.e. $A_{V,\rm stars}= A_{V, \rm gas}$). 

Recently, the possibility of a different dust calibration for high-redshift sources was discussed by \citet{Dayal2012}, \citet{Wilkins2012,Wilkins2013}, \citet{deBarros2014}  and \citet{Castellano2014}. These authors argue that high redshift sources likely have lower metallicities and younger ages than the local starburst galaxies used in the empirical \citet{Meurer1999} calibration. The \citet{Meurer1999} calibration implicitly assumes a dust-free UV-continuum slope of the galaxy of $\beta_{\rm int}=-2.23$, consistent with solar metallicity and ages of a few hundred Myr. However, our $z\sim4$ UV selected galaxies likely have ages around $\sim100$ Myr or less (see \S \ref{sec:SED}), while their metallicity content is expected to be no higher than $0.1-0.5\,\rm Z_\odot$, measured for $z\sim3$ UV selected galaxies by e.g., \citet{Maiolino2008}, \citet{Manucci2009} and \citet{Troncoso2014}. These physical properties result in bluer intrinsic UV-continuum slopes and therefore the dust reddening of the UV-continuum slope could be underestimated when assuming the \citet{Meurer1999} calibration. 

Following the arguments above, the systemic offset between UV and H$\alpha$ based SFR indicators in our $z\sim4$ sample could be explained by an underestimate of our dust-correction due to the implicit assumption of $\beta_{\rm int}=-2.23$. We investigate this possibility by considering a general dust correction of  
\begin{equation}
\label{Eq:dustcorr}
A_{1600} = 1.99\, (\beta-\beta_{\rm int}).
\end{equation}
We vary the intrinsic UV-continuum slope, $\beta_{\rm int}$, to recover a median $\rm SFR_{\rm H\alpha}/SFR_{\rm UV+\beta}\sim1$ after dust correction. Using this method we derive  $\beta_{\rm int}=-2.50^{+0.15}_{-0.06}$ for our spectroscopic and $\beta_{\rm int}=-2.42^{+0.02}_{-0.06}$ for our photometric redshift sample with the bootstrapping uncertainties. These values correspond to an age of $\sim$80 Myr, given a constant star-formation history and metallicity $Z=0.5Z_\odot$. This is somewhat redder than the intrinsic slope of  $\beta_{\rm int}\sim-2.67$ found by \citet{Castellano2014}, but similar to the range of $\beta_{\rm int}$ found by \citet{deBarros2014}. 
 Both of these are derived from SED fitting. 
 Furthermore simulations by \citet{Dayal2012} and \citet{Wilkins2012,Wilkins2013} suggest $\beta_{\rm int}\sim-2.4$, slightly redder than our derived value. 

\begin{figure*}
\centering
\includegraphics[width=0.85\textwidth,trim=0mm 10mm 0mm 0mm] {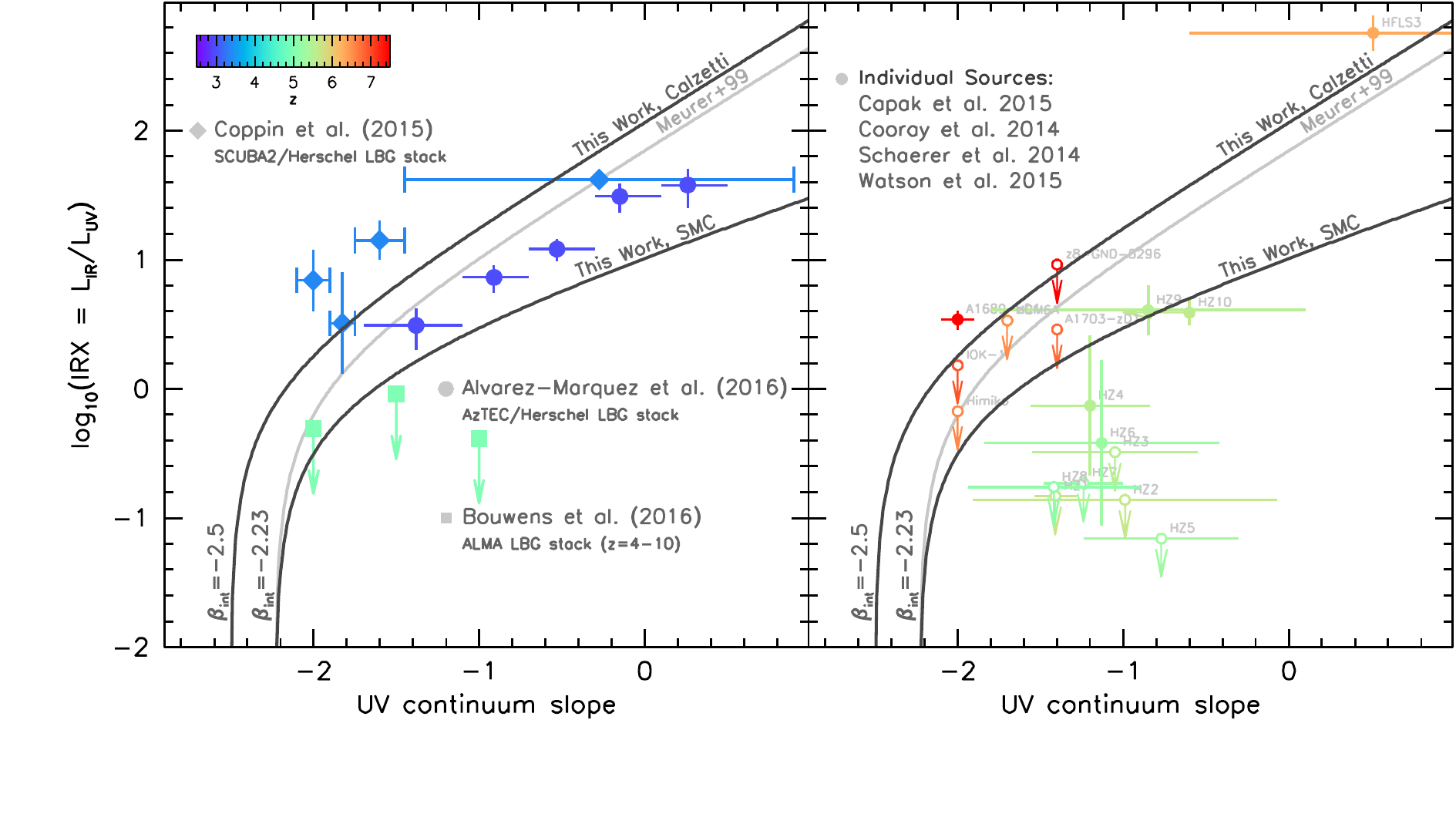} 
\caption{ Current observational constraints on the IR over UV
  luminosity ratio as a function of the UV continuum slope ($\beta$)
  using stacks of UV-selected galaxies (\textit{left panel}) and
  constraints on individual galaxies from ALMA and PdBI (\textit{right
    panel}). While the stacking results by \citet{Coppin2015} over the
  redshift range $z\sim3-5$ are in good agreement with the dust
  calibration needed to bring H$\alpha$ and UV based SFR measurements,
  i.e.  $A_{1600} = 1.99\, (\beta+2.5)$ (see \S
  \ref{sec:origin_dust}), many of the results
  suggest lower dust corrections \citep{Cooray2014,Capak2015,Schaerer2015,Watson2015,AlvarezMarquez2016,Bouwens2016}.  
  At present, the impact that dust
  has on the observed UV brightness and SFRs of $z\sim4$-5 galaxies is
  not clear, on the basis of far-IR observations.}
\vspace*{1mm}
\label{fig:IRX_beta}
\end{figure*}

To assess possible scenarios where the high H$\alpha$ SFRs with respect to UV based SFRs are a result of bluer intrinsic UV-continuum slopes compared to the \citet{Meurer1999} relation, we compare our derived dust correction above with new far-infrared (FIR) and submillimetre constraints from the $Herschel$ Space Telescope, Sub-millimeter Common-User Bolometric Array 2 (SCUBA2), Atacama Large Millimetre Array (ALMA) and Plateau de Bure Interferometer (PdBI) in Figure \ref{fig:IRX_beta}. While stacking analysis of $z\sim2$ galaxy populations generally find good agreement with the \citet{Meurer1999} relation \citep[e.g.][]{Reddy2012,Oteo2014,Pannella2015}, recent measurements at even higher redshifts have reported conflicting results.  \citet{Coppin2015} stack large samples of UV selected galaxies at $z\sim3-5$ using low resolution data from $Herschel$ and SCUBA2 and find consistent high dust content in the typical galaxy (left panel of Figure \ref{fig:IRX_beta}). Furthermore, \citet{Bourne2016} find a slope of 1.8 (as opposed to 1.99) and a $\beta_{\rm int}\sim-2.22$ using stacked \textit{Herschel} and SCUBA-2 imaging of galaxies 
$10^{9}-10^{10}\,M_\odot$ and $z=0.5-6$, closely resembling the \citet{Meurer1999} relationship.

On the other hand the first small samples of high-resolution individual dust continuum detections and constraints on UV-selected $z\gtrsim4$ galaxies \citep[e.g.][]{Capak2015,Schaerer2015,Knudsen2016} indicate a typical dust content significantly below the \citet{Meurer1999} relation (right panel of Figure \ref{fig:IRX_beta}). Indeed, recent stacking analysis of ALMA observations of low-mass galaxies in the Hubble Ultra Deep Field \citep{Aravena2016,Bouwens2016,Dunlop2016} indicate sources below $10^{10}M_\odot$ show surprisingly faint infrared luminosities. In particular, comparing the low  infrared excess as a function the UV-continuum slope in a sample of stacked galaxies below a stellar mass of $10^{9.75}M_\odot$, \citet{Bouwens2015} find that the slope of the IRX-$\beta$ relation ship has to be $<1.22$ and $<0.97$ for $z\sim2-3$ and $z\sim4-10$ galaxies respectively (Fig. \ref{fig:IRX_beta})

In conclusion, the first high-resolution FIR/submm observations taken by ALMA  would indicate that the dust attenuation law is closer to that found for the Small Magellanic Cloud \citep[SMC; e.g.][]{Prevot1984}. \citet{Oesch2013} show that an SMC-type dust-law is also preferred by the relationship between the UV-continuum slope and the UV-to-optical color of $z\sim4$ galaxies. 

\begin{figure*}
\centering
\includegraphics[width=1.\textwidth,trim=17mm 175mm 33mm 35mm] {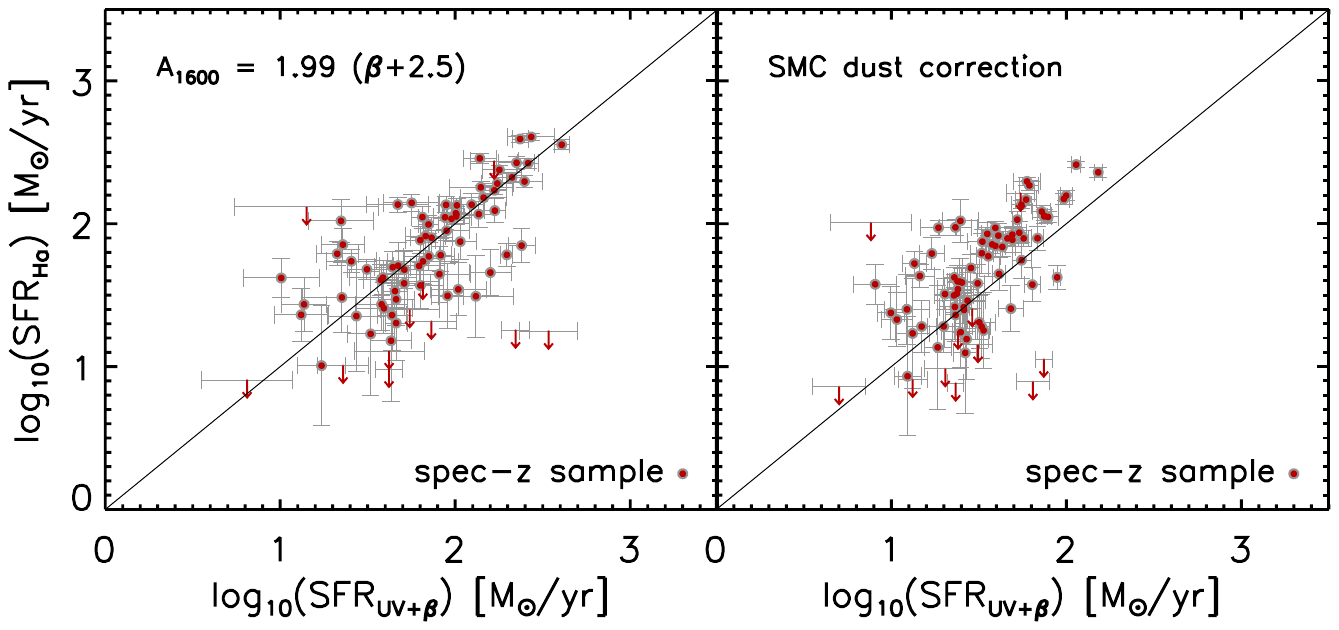} 
\caption{The SFR$_{\rm H\alpha}$ as a function of SFR$_{\rm UV+\beta}$, using a dust-correction calibrated to reproduce  a median $\rm SFR_{\rm H\alpha}/SFR_{\rm UV}\sim1$ (\textit{left panel}, see  \S \ref{sec:origin_dust}) and the same figure assuming an SMC-type dust correction (\textit{right panel}) for our spectroscopic sample; H$\alpha$ is corrected for dust assuming $A_{V,\rm stars}= A_{V, \rm gas}$ and using the \citet{Calzetti2000} and SMC \citep{Prevot1984} dust curve in the left and right panel respectively.
The direct calibration of different SFRs at $z\sim4$ will be improved in the near future with large samples of galaxies with sensitive high-resolution dust continuum measurements from ALMA or PdBI. 
} 
\label{fig:SFR_newcal}
\vspace*{2mm}
\end{figure*}

The ambiguity in the dust-correction of typical high-redshift UV-selected galaxies outlined in Figure \ref{fig:IRX_beta} has significant implications for the interpretation of the high H$\alpha$ EWs and H$\alpha$ SFRs derived in this work. In Figure \ref{fig:SFR_newcal} we show the H$\alpha$ and UV based SFRs, when applying two different dust corrections. The first (\textit{left panel}) adopts a $A_{1600} = 1.99\, (\beta+2.5)$ prescription.  By construction, this produces an excellent match between H$\alpha$ and UV based SFRs (and assumes an intrinsic UV-continuum slope $\beta_{\rm int}=-2.5$).  In this case, the high-EW nebular emission lines found in high-redshift galaxies are in part the result of relatively high levels of dust-obscured star formation in combination with a similar dust attenuation between stars and the nebular light.  

The right panel in Figure \ref{fig:SFR_newcal} uses 
\begin{equation}
\label{Eq:SMCdust}
A_{1600} = 1.1\, (\beta+2.23),
\end{equation}
 corresponding to a SMC-type dust-correction for an intrinsic
 UV-continuum slope $\beta_{\rm int}=-2.23$, similar as that
 implicitly assumed for the \citet{Meurer1999} relation. We find a
 linear slope 1.1 from the tabulated extinction values of the SMC by
 \citet{Prevot1984}. The resulting low levels of dust
 obscuration imply that, even for similar levels of attenuation
 between nebular and stellar light, the H$\alpha$ SFRs are $0.19^{+0.02}_{-0.02}$ dex 
  in the spectroscopic and $0.17^{+0.02}_{-0.03}$  dex  in the photometric redshift sample
above what we would expect for the total amount of star-formation
 derived from the dust-corrected UV light.  While the samples of
 individual galaxies with high resolution, high signal-to-noise constraints on
 the FIR light are still small, these results have significant
 consequences if correct.  This provides us with the motivation to explore
 other mechanisms for recovering high $\rm SFR_{\rm H\alpha}/SFR_{\rm
   UV+\beta}$ ratios.

\begin{figure*}
\centering
\includegraphics[width=1.\textwidth,trim=17mm 175mm 33mm 35mm] {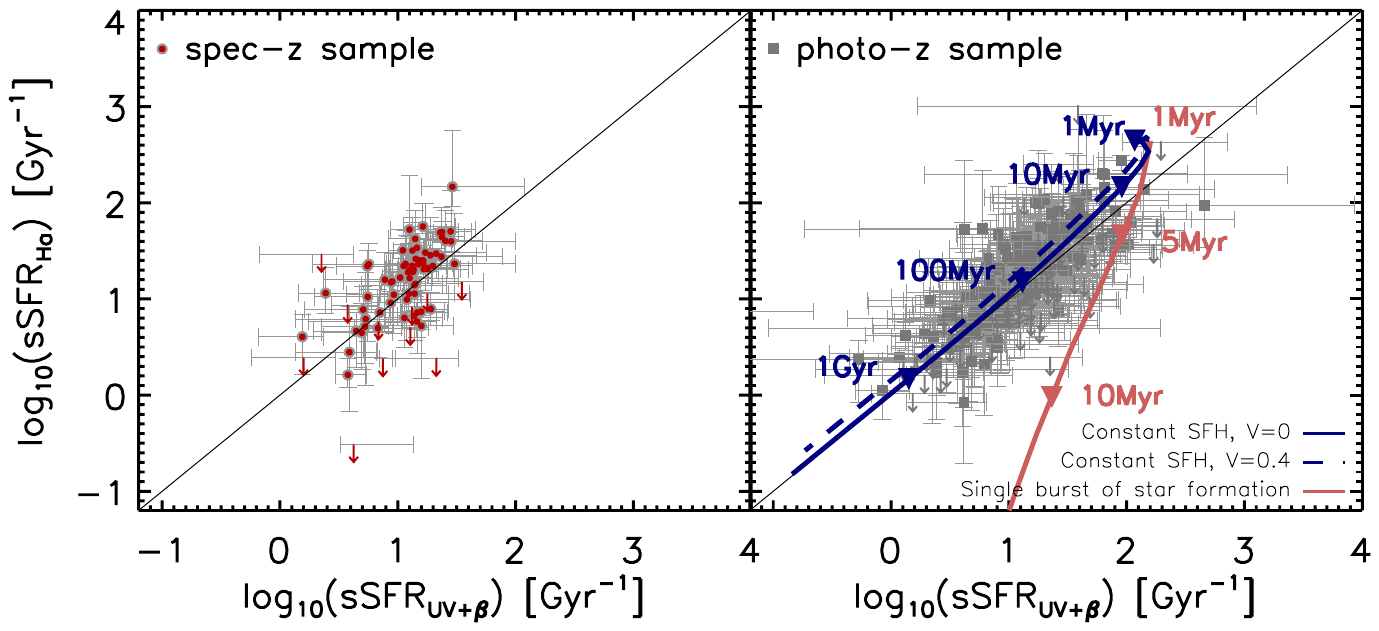} 
\caption{The specific star formation rates from the inferred H$\alpha$ luminosities versus those from the UV-luminosity corrected for dust using the UV slope $\beta$ and the \citet{Meurer1999} calibration (red points; red arrows indicate the 1$\sigma$ upper limits); H$\alpha$ is corrected for dust assuming $A_{V,\rm stars}= A_{V, \rm gas}$ and using the \citet{Calzetti2000} dust curve.   The left panel shows our spectroscopic sample, while the right panel shows our photometric sample. The right panel includes two stellar population models from Starburst99 \citet{Leitherer1999}. The rose color track includes only single star populations and follows a single burst of star formation with an initial mass of $10^6\,M_\odot$. We indicate the measured sSFRs of this model using the \citet{Kennicutt1998} relations at ages of 1 Myr, 5 Myr and 10 Myr (rose triangles). The dark blue tracks indicate stellar populations with a constant star formation history of $1\,M_\odot\rm  yr^{-1}$, where the solid line indicates a stellar population with no rotation (V=0) and the dashed line indicates a stellar population with rotation levels at 40\% of the break-up velocity \citep[V=0.4][]{Leitherer2014}. We indicate the measured sSFRs of this model using the \citet{Kennicutt1998} relations at ages of 1 Myr, 10 Myr, 100 Myr and 1 Gyr (dark blue triangles). The one-to-one relation in the sSFRs favors a relatively smooth star-formation history, given the fact that very few sources are at low sSFR$_{\rm H\alpha}$ as would be the case for a starburst with an age of $>$10 Myr.
} 
\vspace*{1mm}
\label{fig:sSFR_comp}
\end{figure*}

\begin{figure*}
\centering
\includegraphics[width=1.\textwidth,trim=17mm 175mm 33mm 35mm] {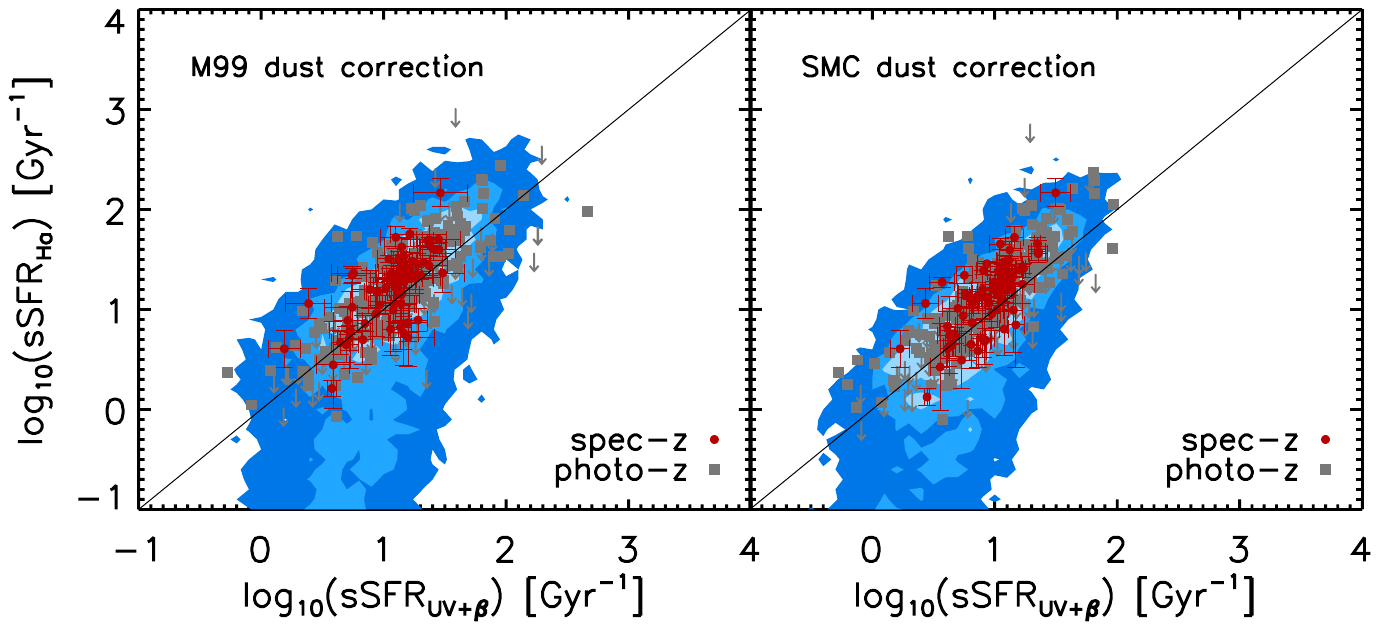} 
\caption{The specific star formation rates from the inferred H$\alpha$ luminosities versus those from the UV-luminosity corrected for dust using the UV slope $\beta$ and the \citet{Meurer1999} calibration (\textit{left panel}) or a SMC-type calibrations (\textit{right panel}); H$\alpha$ is corrected for dust assuming $A_{V,\rm stars}= A_{V, \rm gas}$ and using the \citet{Calzetti2000} and SMC \citep{Prevot1984} dust curve in the left and right panel respectively. Data points indicate the observed spectroscopic (red points) and photometric redshift selected (gray squares) samples. The blue contours show a simulated galaxy distribution with largely similar masses and specific star-formation rates as our observed galaxy samples, assuming a bursty star-formation history with burst masses $M_{\rm burst}\sim10^8 M_\odot$ and burst intervals of $d \rm t_{burst}\sim5-10 Myr$ (see \S\ref{sec:origin_bursty}). The simulations are cut below $\rm SFR_{\rm UV} < 5 \, M_\odot yr^{-1}$ to mirror the UV-selection of the observed sample. Post-starburst galaxies are visible in the UV for $\sim100$ Myr while H$\alpha$ probes the instantaneous SFR. The fraction of sources in the resulting low $\rm sSFR_{\rm H\alpha}$ tail is twice as large in the simulated distribution as found in the observed sample (including upper limits).
} 
\vspace*{1mm}
\label{fig:bursts}
\end{figure*}

\subsection{Star-formation histories}
\label{sec:SFH}

In the previous section we investigated the effect of different dust calibrations on the  $\rm SFR_{\rm H\alpha}/SFR_{\rm UV+\beta}$ ratio of $z\sim4$ UV-selected galaxy samples. While it is possible to reconcile H$\alpha$ and UV based SFR estimates, using the \citet{Kennicutt1998} relations and assuming $A_{V,\rm stars}= A_{V, \rm gas}$ and dust corrections slightly above what is predicted by the \citet{Meurer1999} relation, it is not clear that such a dust law is supported by the observations of $z\geq4$ galaxies.  

The first direct high-resolution dust-continuum measurements of
UV-selected high-redshift galaxies suggest that the typical dust
content of these galaxies is significantly lower than this (see figure
\ref{fig:IRX_beta}).  If future observations give similar results --
implying more of an SMC dust law -- our estimated SFRs using H$\alpha$
would be systematically higher than using the $UV$ light.  Here, we
will investigate how different star-formation histories can play a
role in the derived SFRs when using the locally derived
\citet{Kennicutt1998} relations.

 \subsubsection{Bursty star formation histories}
 \label{sec:origin_bursty}
One possible explanation for the high $\rm SFR_{\rm H\alpha}/SFR_{\rm
  UV+\beta}$ ratios we derive in \S \ref{sec:SFRs} would be that the
galaxies we examine are all predominantly young, i.e., such that the
H$\alpha$-based SFRs are much larger than $UV$-based SFRs (which
saturate at $\sim$100 Myr).  However, it is possible that these
galaxies are much younger, or perhaps that galaxies undergo regular
bursts of star formation \citep[e.g.][]{Dominguez2015} which would 
give them the appearance of very
young systems. This would boost the H$\alpha$ flux, which is
predominantly generated by short-lived ($<$10 Myr) O-stars, with
respect the observed UV light, which is produced by O- and B-stars on
a somewhat longer timescale ($\sim$100 Myr).  This is the hypothesis
that \citet{Shim2011} favour for explaining the high $\rm SFR_{\rm
  H\alpha}/SFR_{\rm UV}$ in the sample they observe.
 
In Figure \ref{fig:SFR_comp} we show SFR$_{\rm H\alpha}$ as a function of SFR$_{\rm UV+\beta}$ for both our spectroscopic sample (left) and our photometric sample (right).  Since the majority of the sources in the left panel show Ly$\alpha$ in emission, one might suppose this population of sources could be biased towards starbursting systems relative to samples which are photometric-redshift-selected.  However, we find little difference in the median $\rm SFR_{\rm H\alpha}/SFR_{\rm UV+\beta}$ ratio of our spectroscopic and our photometric sample. Moreover, a subsample of sources with stellar masses above our mass completeness limit of $\sim 10^{10}\,\rm M_\odot$ gives a comparable median $\rm SFR_{\rm H\alpha}/SFR_{\rm UV+\beta}$ ratio to that of  the entire photometric sample. These findings argue against supposing that very young ($<$10 Myr) starburst ages drive the systematic offsets we observe between H$\alpha$ and UV SFRs. 

To gain further insight we show the \textit{specific} SFR$_{\rm H\alpha}$ as a function of the specific SFR$_{\rm UV+\beta}$ in Figure \ref{fig:sSFR_comp}, again assuming a \citet{Meurer1999} dust law and $A_{V,\rm stars}= A_{V, \rm gas}$. The two sSFR estimates correlate strongly over $\sim$2 dex in sSFR and we find a constant offset between the H$\alpha$ and UV probes of $\sim0.15$ dex. For young galaxies that are formed in a single burst of star-formation we would expect the discrepancy between sSFR$_{\rm H\alpha}$ and  sSFR$_{\rm UV+\beta}$ to decrease with decreasing sSFR$_{\rm H\alpha}$. For reference we include a single stellar population (SSP) track (rose line) that demonstrates the rapid evolution of the $\rm sSFR_{\rm H\alpha}/sSFR_{\rm UV+\beta}$ ratio of this stellar population with age. A burst of star formation is expected to show enhanced H$\alpha$ for $\sim$5 Myr, but we would expect many sources with ages $>$10 Myr after the burst below the one to one relation. Comparing this model with our observations, we see no clear trend in favor of young ages.

To test this scenario further we run a Monte-Carlo simulation, which we consider a population of galaxies with bursty star-formation histories.  To reproduce the properties in our sample we randomly draw from the distribution of derived stellar masses for our observed galaxy sample and populate each galaxy with bursts of mass $M_{\rm burst}$ distributed linearly in time with a typical time interval $\rm dt_{burst}$. For each burst we add a SSP model obtained from the Starburst99 models \citep{Leitherer1999} with the corresponding age to the total spectral energy distribution of the simulated source. We derive $\rm SFR_{UV}$ and $\rm SFR_{H\alpha}$ from the final SED using the \citet{Kennicutt1998} relations. Because our observed galaxy sample is limited by the $H_{160}$ band flux, we assume that the galaxy population can be modeled as an SFR$_{\rm UV}$-limited sample (assuming SFR$_{\rm UV}$ scales linearly with $L_{\rm UV}$ through Eq. \ref{Eq:Kennicutt_UV}). Therefore, we impose a $\rm SFR_{UV}$ lower-limit on our simulated galaxy population of 5 $M_\odot \rm yr^{-1}$. We assign observational errors to the simulated datapoints, using the observed uncertainties in the derived SFRs. Using these simulations we investigate what parameters of $M_{\rm burst}$ and $\rm dt_{burst}$ can roughly reproduce a galaxy population with a similarly high $\rm SFR_{\rm H\alpha}/SFR_{\rm UV+\beta}$  ratio when assuming the \citet{Kennicutt1998} relations as the typical source in our observed galaxy samples. 

Figure \ref{fig:bursts} shows two such simulations that produce a large number of sources with $\rm SFR_{\rm H\alpha}/SFR_{\rm UV+\beta}>1$ . To reproduce the observed galaxy distribution we find that we typically need high burst masses of $M_{\rm burst}\sim10^8 M_\odot$, to reproduce the generally high sSFRs, and reasonably short burst intervals of $\rm dt_{burst}\sim5-10 Myr$ that generate high H$\alpha$ fluxes. 
While roughly half of the galaxies in our simulation have $\rm SFR_{\rm H\alpha}/SFR_{\rm UV+\beta}>1$, our simulations also show a rather large tail of relatively low H$\alpha$ sSFR galaxies, i.e. $\sim53\%$ of the simulated galaxy distributions in the left panel of Figure \ref{fig:bursts} have $\rm sSFR_{\rm H\alpha}<2\,  yr^{-1}$. This is in contrast to the observed galaxies in our large photometric sample where $\sim26\%$ of the sources have upper limits in $\rm SFR_{\rm H\alpha}$ and could therefore populate this low-sSFR tail.

In conclusion, we find that bursty star-formation histories predict
at least twice as many galaxies with $\rm sSFR_{\rm H\alpha}<2\, yr^{-1}$ as
are seen in the observations.  The implication is that the star
formation histories of galaxies are considerably more smooth than in
the toy model we consider above and that bursty star-formation
histories do not provide a resolution for the tension between the
H$\alpha$ and UV-based SFRs. These results are in agreement with 
simulated star-formation histories which predict that burstiness is
mostly present in low mass ($\lesssim 10^8 M_\odot$) galaxies \citep[e.g.][]{Dayal2013,Sparre2015}.

\subsubsection{Rising star formation histories}
\label{sec:origin_rising}

In \S \ref{sec:origin_bursty} we describe how the distribution of
H$\alpha$ and UV-based sSFRs disfavors bursty star formation
histories. However, a smoothly rising star-formation history can also
affect the $\rm SFR_{\rm H\alpha}/SFR_{\rm UV+\beta}$ ratio simply
because the UV flux probes the time-averaged SFR over a $\sim$100 Myr
time window.  As the SFR for rising star-formation histories is lower
at earlier times in a $\sim$100 Myr time window, the SFR inferred from
the $UV$ light would be lower than the instantaneous SFR.

\citet{Reddy2012} tabulate the values of SFR/L$_{1700}$ as a function
of galaxy age for different star formation histories (see their Table
6). Using their tabulated values, an exponentially rising star
formation history ($\tau\sim100$) results in a $\sim0.07$ dex higher
SFR/L$_{1700}$ ratio for a galaxy of 100 Myr compared to a constant
star-formation history. Assuming H$\alpha$ is a good tracer of the
instantaneous SFR we estimate that rising star formation histories can
reasonably result in a $\sim0.1$ dex offset $\rm SFR_{\rm
  H\alpha}/SFR_{\rm UV+\beta}$ ratio.

Similar to the scenario of bursty star-formation histories, rising star-formation histories can work well in combination with a \citet{Meurer1999} dust correction to explain the values of $\rm SFR_{\rm H\alpha}$ and $\rm SFR_{\rm UV+\beta}$ in our $z\sim4$ galaxy sample. However, if these galaxies prefer a SMC-type dust correction such as suggested by e.g. \citet{Capak2015}, one cannot explain the offset in the derived $\rm SFR_{\rm H\alpha}/SFR_{\rm UV+\beta}$ values just invoking rising star-formation histories.

\begin{table*}
\begin{threeparttable}
\centering
\caption{Observationally-Motivated Physical Assumptions (see Table~\ref{tab:summary}) Used in Deriving our Fiducial $z=4$-8 SFR Functions and Star-Formation Main-Sequence Results at $z\sim4.3$}
\begin{tabular}{lccc} 
\hline 
\hline
Assumptions SFR functions & $\log_{10}(\rm SFR_{\rm H\alpha}/SFR_{\rm UV})^a$ & $\Delta\log_{10}(\rm SFR_{H\alpha})^b$\\
\hline
Fiducial Model  & $-0.01^{+0.02}_{-0.02}$ & $-0.20$\\
SMC dust correction, $A_{V,\rm stars}= A_{V, \rm gas}$\\
Stellar rotation (see \S\ref{sec:origin_ionrad} and Table~\ref{tab:summary})\\
Rising SFH (see \S\ref{sec:origin_rising} and Table~\ref{tab:summary})\\
(results presented in \S\ref{sec:implications}: Figures~\ref{fig:mainseq}-\ref{fig:SFRfunc} and 
Tables~\ref{tab:sfr_mass}-\ref{tab:stepwise_sfr_SMC})\\
\hline
Alternate Fiducial Model & $0.09^{+0.02}_{-0.04}$ & $+0.12$\\
Meurer+99, $A_{V,\rm stars}= A_{V, \rm gas}$ \\
Rising SFH (see \S\ref{sec:origin_rising} and Table~\ref{tab:summary})\\
(results presented in appendix B: Tables~\ref{tab:schechter_sfr}-\ref{tab:stepwise_sfr})\\
\hline
\end{tabular} 
\label{tab:fiducial}
$^a$ Median ratio of the SFRs derived from H$\alpha$ and the $UV$-continuum light making use of the assumptions in these observationally-motivated physical models.\\
$^b$ Value of the Inferred SFRs using the Fiducial Assumptions minus the SFRs derived using the \citet{Kennicutt1998} relations with no dust corrections.  SFRs calculated using our fiducial model assuming SMC extinction will be systematically 0.32 dex lower than in our alternate fiducial model with \citet{Meurer1999} extinction.\\
\end{threeparttable}
\end{table*}

\subsection{Production Efficiency of Ionizing Photons}
\label{sec:origin_ionrad}

In \S \ref{sec:SFH} we discussed the impact of the assumed star
formation history on the offset derived from H$\alpha$ and UV-based
SFR indicators in our $z\sim4$ galaxy sample. While bursty or rising
star formation histories will produce a $\sim$0.1-dex systematic
offset between the two SFR indicators, this does not resolve the
tension between these two SFR measures adopting an SMC dust law (where
there is a $\sim$0.2-dex offset: see Figure \ref{fig:SFR_newcal} and
Table~\ref{tab:summary}).

Another effect on the $\rm SFR_{\rm H\alpha}/SFR_{\rm UV+\beta}$ ratio
that we must consider is the potential for a changing conversion
factor between $L_{\rm H\alpha}$ and SFR
(i.e. Eq. \ref{Eq:Kennicutt_Ha}, see also \citealt{Zeimann2014}). While the H$\alpha$ flux scales
directly with the number of ionizing photons emerging from the
\ion{H}{2} regions in the galaxy \citep{Kennicutt1998,Leitherer1995},
the shape of the ionizing spectrum in low metallicity galaxies is
poorly constrained. In particular the impact of massive binaries \citep[e.g.][]{deMink2009,Sana2012},
rotational mixing \citep[e.g.][]{Ramirez2013} and line blanketing can
change with metallicity. For a more elaborate discussion we refer to
\S3 in \citet{Kewley2013} and the discussion in \citet{Steidel2014}. 
Furthermore, differences in the high-mass
slope of the IMF would also introduce a different ionizing spectrum.
Even without changes in the IMF a metallicity dependent ionizing
spectrum could significantly impact the $\rm SFR_{\rm
  H\alpha}/SFR_{\rm UV+\beta}$ ratio.  If stars in high-redshift
galaxies are really much more efficient producers of ionizing photons,
this would significantly impact galaxies' possible role in reionizing
the universe \citep{Bouwens2015,Stark2015,Stark2016}.

We illustrate this in the right panel Figure \ref{fig:sSFR_comp}, where we show stellar population tracks for a constant star-formation history for stars with zero rotation (at solar metallicity) and for stars that rotate at 40\% of the break-up velocity (at $Z=0.6Z_\odot$) from the models described in \citet{Leitherer2014}. At 100 Myr, the low metallicity models that includes stellar rotation are $\sim0.15$ dex offset in the $\rm SFR_{\rm H\alpha}/SFR_{\rm UV+\beta}$ ratio compared to the model that does not include stellar rotation.

\begin{figure*}
\centering
\includegraphics[width=1.02\columnwidth,trim=115mm 175mm 25mm 38mm] {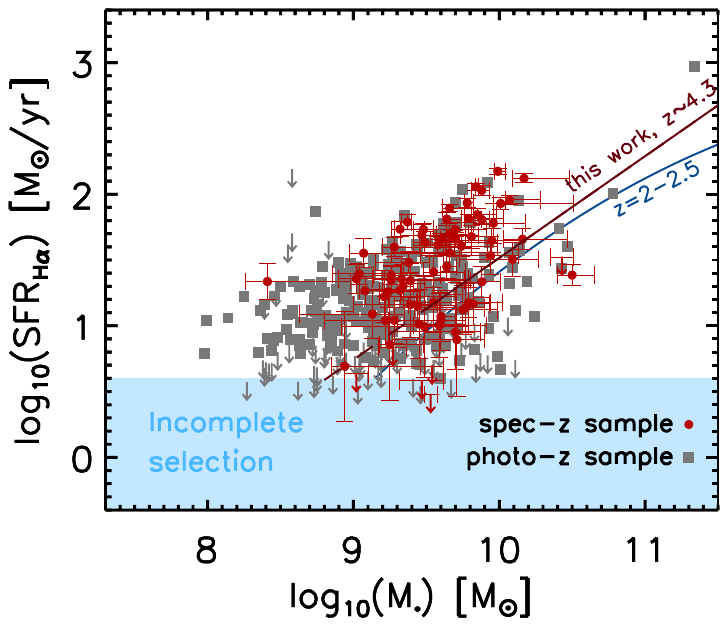} 
\includegraphics[width=1.02\columnwidth,trim=30mm 175mm 110mm 38mm] {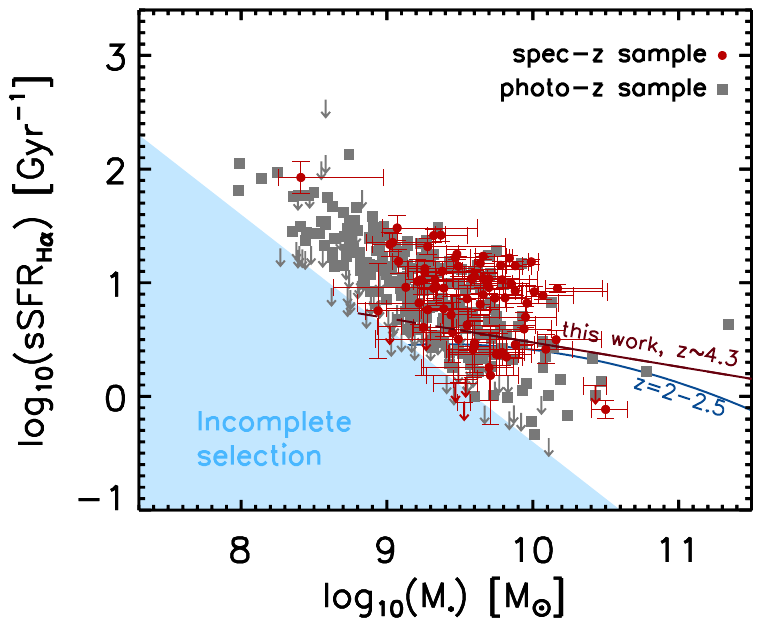} 
\caption{The SFR$_{\rm H\alpha}$ (left) and sSFR$_{\rm H\alpha}$ (right) as a function of stellar mass for our spectroscopic (red points) and photometric (grey squares) sample respectively. The  solid red lines indicates the Bayesian linear regression for galaxies $M_\ast > 10^{9.5}M_\odot$, while the blue line indicates the polynomial derived by \citet{Whitaker2014} for $z=2.0-2.5$ galaxies. The blue shaded region gives an indication of the incompleteness in our sample due to the UV selection. We find that the slope of the SFR-stellar mass sequence is broadly consistent with unity and an intrinsic scatter of $\lesssim0.4$ dex.} 
\label{fig:mainseq}
\end{figure*}

A similar effect is seen when using the  \citet{Eldridge2012} models that include binary star evolution. Comparing their models that include binaries at $Z=0.2Z_\odot$ with the model for single star evolution at solar metallicity we find a $\sim0.31$ dex offset in the $\rm SFR_{\rm H\alpha}/SFR_{\rm UV+\beta}$ ratio, arising from a $\sim0.46$ dex offset in H$\alpha$ flux and $\sim0.15$ dex in UV luminosity. 


Possible evolution in the SFR-$L_{H\alpha}$ relationship therefore
offers us a way of explaining the observed high H$\alpha$ fluxes, even
in the scenario where the typical high-refshift UV-selected galaxy has
low dust masses such as been argued by \citet{Schaerer2015} and
\citet{Capak2015}. 


In Table \ref{tab:summary} we give an overview of how each of the
models considered in this section impacts the $\rm SFR_{\rm
  H\alpha}/SFR_{\rm UV+\beta}$ ratios.  Over the next few years, ALMA
will likely shed light on the typical dust properties of $z\sim4$
galaxies, and as a result provide us with new insights into the
star-formation histories and ionizing spectra of high-redshift
galaxies.

\section{Implications}
\label{sec:implications}

In the previous section, we considered a variety of different physical
mechanisms for reconciling current measures of the SFRs as derived
from H$\alpha$ or from $UV$-continuum light.

On the basis of this discussion (and comparison with the
observations), we find that there are at least two flavors of physical
models that appear plausible.  The first supposes that $z\sim4$
galaxies can be described using a \citet{Meurer1999} dust calibration
with a $A_{V,\rm stars}= A_{V, \rm gas}$ and that the $UV$-based SFR
estimates need to be corrected by $\sim$0.1 dex to correct the
measured, time-averaged values to the instantaneous ones.

The second supposes that $z\sim4$ galaxies can be described using a
SMC dust calibration with a $A_{V,\rm stars}= A_{V, \rm gas}$ and that
$z\sim4$ galaxies are more efficient at producing ionizing photons
than in standard stellar population models (and thus the
$L_{H\alpha}$/SFR ratio is high: see \citealt{Bouwens2015,Stark2015,Stark2016} for a
discussion of how this may impact galaxies' role in driving the
reionization of the universe).

These two scenarios are summarized in Table~\ref{tab:fiducial}.  Which
of these scenarios is the relevant one largely hinges on the dust law (see
Figure \ref{fig:IRX_beta}) and should be resolved definitively in the near future
with deeper ALMA data and larger samples of indiviual detections of
 high-redshift galaxies. 
 Given that the current ALMA results of both indiviually detected UV-bright galaxies 
 \citep[e.g.,][]{Capak2015} and large samples of faint stacked galaxies \citep[e.g.,][]{Bouwens2016}  indicate a signficantly  lower infrared excess for  low-mass high-redshift  than predicted by the \citet{Meurer1999} calibration, we will assume for the remainder of this section that the latter
scenario, assuming a SMC-type dust calibration, provides a reasonable 
basis from which to derive new results
on the SFR-stellar mass relation and also the $z=4$-8 SFR functions.
However, we will also look at the results if the former scenario
involving the \citet{Meurer1999} dust law is the correct one in appendix B.

\subsection{SFR-stellar mass sequence}
\label{sec:mainseq}
One of the most fundamental relations for understanding galaxy
build-up is the SFR-stellar mass relation, or the ``main sequence'' of
star-forming galaxies. Using our derived H$\alpha$-based SFRs we are
in an excellent position to assess this relation at $z\sim4$, given
the much weaker sensitivity of our H$\alpha$-based SFR measurements to
many of the classic degeneracies that affect stellar population
modelling (e.g. dust vs. age).  For this analysis we will make use of
the H$\alpha$ measurements corrected as described in \S
\ref{sec:sfrHa} to obtain a good estimate of the instantaneous SFR.
As specified at the beginning of this section, we utilize an SMC dust 
correction (Eq. \ref{Eq:SMCdust}) and $A_{V,\rm stars}= A_{V, \rm
  gas}$. As a result we have to conclude that a higher production 
  efficiency of ionizing photons impact the  $\rm SFR_{\rm H\alpha}/SFR_{\rm UV+\beta}$ 
  ratio and therefore we correct the H$\alpha$ SFRs downwards by $0.2$ dex 
  compared to the values obtained with the \citet{Kennicutt1998} relation (Eq. \ref{Eq:Kennicutt_Ha}). 
   See Table~\ref{tab:summary} to see what impact other dust
corrections would make to our final result.

In Figure \ref{fig:mainseq} we show SFR$_{\rm H\alpha}$ and sSFR$_{\rm
  H\alpha}$ as a function of stellar mass for our spectroscopic and
photometric samples.  Since only sources with $H_{160,AB}$ magnitudes
brighter than 26.5 (see \S\ref{sec:phot_sample}) were included in our
sample (equivalent to a SFR$_{\rm UV}$ limit of
$\sim4\,M_\odot\rm yr^{-1}$), we present this selection limit very
clearly on this figure.
We find that a fit using only sources above $10^{9.8} M_\odot$  is consistent within the uncertainty with the unity low-mass slope as found by \citet{Whitaker2014} for star-forming galaxies between $z\sim0.5$ and 2.5. 

Furthermore we estimate the scatter in the main sequence of star-forming galaxies from the Bayesian linear regression (solid line in the left panel of Figure \ref{fig:mainseq}) with a flat prior \citep{Kelly2007}, which gives an intrinsic scatter of $\sim0.4$ dex, indicative of a modestly smooth star formation history. This intrinsic scatter is significantly higher than the $\sim0.13$ dex scatter  measured by \citet{Speagle2014} based on the \citet{Shim2011} sample, but in good agreement with the recent determination from \citet{Salmon2015}.

While the dynamic range where we have a mass complete sample is limited, we can compare the normalization of our SFR-stellar mass sequence with determinations at lower redshift in more detail. At a stellar mass of $10^{10}\,M_\odot$ we find from our Bayesian fits $\rm \log_{10} SFR_{\rm H\alpha}/M_\odot\rm yr^{-1}=1.51\pm0.07$ and $\rm \log_{10} sSFR_{\rm H\alpha}/\rm Gyr^{-1}=0.47\pm0.06$ (uncertainties obtained through bootstrapping). This is slightly lower than the fit by \citet{Speagle2014}, who compare 25 studies between $z\sim0$ and $z\sim6$ and predict an SFR of  $\rm \log_{10} SFR/M_\odot\rm yr^{-1}\sim1.73$ (corrected for differences in IMF) at our median redshift $<z_{\rm spec}>=4.25$ and stellar  mass of $10^{10}\,M_\odot$. Furthermore, extrapolating the relation for $\rm sSFR\propto(1+z)^{1.9}$ found by \citet{Whitaker2014} between $z\sim0.5$ and 2.5, we would predict $\rm \log_{10} SFR/M_\odot\rm yr^{-1}\sim2.05$ at $10^{10}\,M_\odot$ if this relation would hold out to $z\sim4$. The lower SFRs in comparison with these lower redshift extrapolation could indicate a flatter evolution of the main-sequence of star-forming galaxies with redshift above $z>2$ such as suggested by \citet{Gonzalez2014} and \citet{Marmol2015}. 

Comparing our normalization of the main sequence with recent estimates at the same redshift we find lower values than \citet{Stark2013} and \citet{Marmol2015} who find $\rm \log_{10} sSFR_{UV+\beta}/\rm Gyr^{-1}\sim0.79$ at $5\cdot 10^{9}\,M_\odot$ and $\rm \log_{10} sSFR_{UV+\beta}/\rm Gyr^{-1}\sim0.73$ at $10^{10}\,M_\odot$ respectively. 
Although these authors use the same technique for deriving the H$\alpha$ SFR, they assume a \citet{Meurer1999} dust correction, which explains the discrepancy.  Our determinations are in good agreement with the results by \citet{Gonzalez2014}, who find $\rm \log_{10} sSFR_{\rm SED}/\rm Gyr^{-1}\sim0.54$ at $5\cdot 10^{9}\,M_\odot$ using sSFRs from SED fitting. 
On the other hand we find slightly higher values than two recent studies using SED fitting results; i.e. \citet{Duncan2014} find $\rm \log_{10} sSFR_{\rm SED}/\rm Gyr^{-1}\sim0.37$ at $5\cdot 10^{9}\,M_\odot$ and \citet{Salmon2015} find a median $\rm \log_{10} SFR_{\rm SED}/M_\odot\rm yr^{-1}\sim1.35$ at $z\sim4$ and $10^{10}\,M_\odot$ (approximately implying $\rm \log_{10} sSFR_{\rm SED}/\rm Gyr^{-1}\sim0.35$). Differences between SFRs derived from SED fitting and H$\alpha$ inferred SFRs could be due to the assumed star-formation history, dust-law and metallicity assumed in both methods, the age-dust degeneracy in the SED fitting and the stellar library that is used for the SED fit; in particular, including binary and/or rotating stars changes the strength of the H$\alpha$ EW at a fixed galaxy age, which influences the SFR and sSFR fitting parameters obtained in the SED fitting. 

We summarize our findings on the SFR$_{\rm H\alpha}-M_\ast$ sequence in Table \ref{tab:sfr_mass}.

\begin{figure*}
\centering
\includegraphics[width=1.65\columnwidth,trim=20mm 163mm 55mm 40mm] {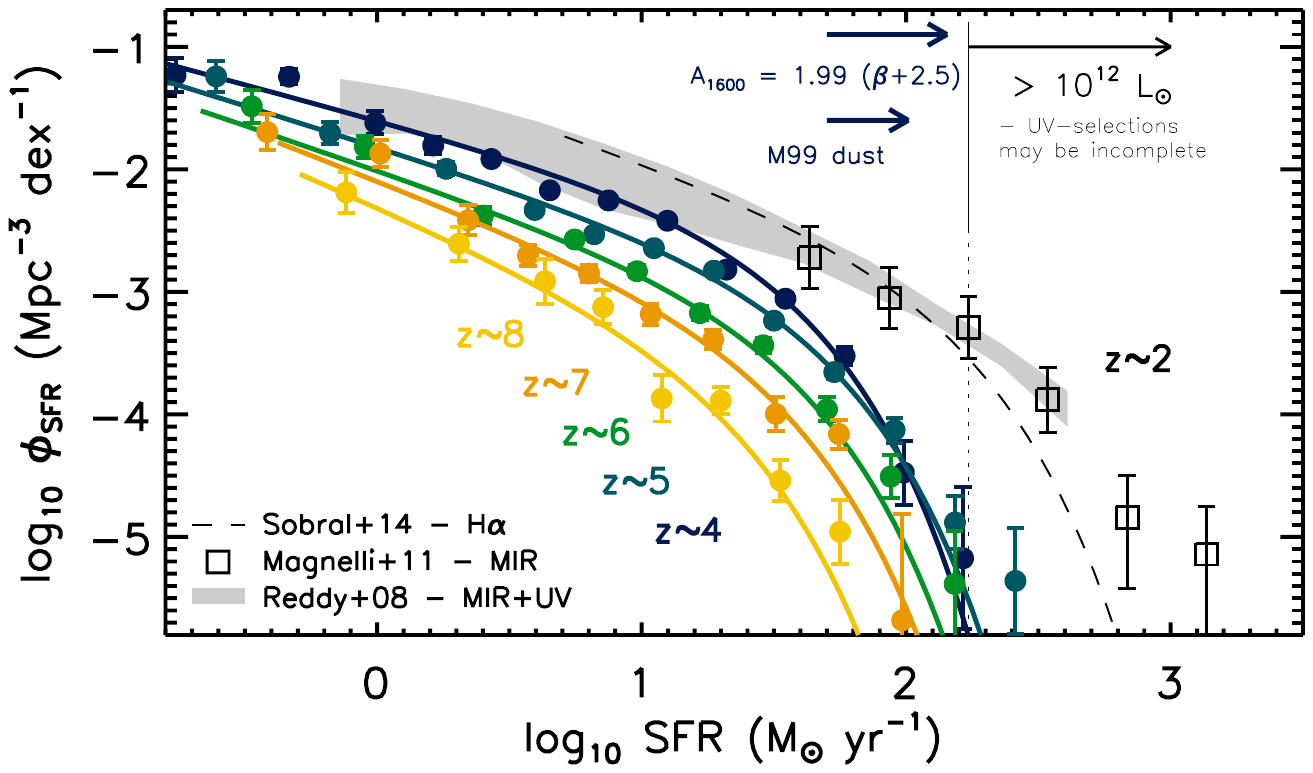} 
\caption{The $z=4$-8 SFR functions derived here following the \citet{Smit2012} procedure. The SFR functions are based on the UV luminosity functions by \citet{Bouwens2014}, the color magnitude relations determined by \citet{Bouwens2013} and the SMC-type dust calibration (Eq. \ref{Eq:SMCdust}). We assume the \citet{Kennicutt1998} conversion from UV to SFR (Eq. \ref{Eq:Kennicutt_UV}).  Stepwise dust-corrected SFR functions (solid points) with the analytical solutions for the Schechter functions \citep[see Eq. 4, 7 and 8 of][]{Smit2012}. The black dotted line indicates the SFR range where the \citet{Bouwens2014} UV selection could be incomplete due to dust saturation or the dust corrections inaccurate. The dark blue arrows indicates the change in the knee of the SFR function assuming different dust corrections $A_{1600}=1.99\,(\beta+2.5)$ or $A_{1600}=1.1\,(\beta+2.23)$ and using the \citet{Kennicutt1998} conversion from UV to SFR (see \S\ref{sec:origin_dust}).  For reference we include SFR functions at $z\sim2$ from H$\alpha$ \citep[dashed black line][]{Sobral2014}, MIR \citep[open black squares][]{Magnelli2011} and UV+MIR \citep[gray shaded region][]{Reddy2008} SFR probes.  } 
\label{fig:SFRfunc}
\vspace*{3mm}
\end{figure*}  
 
\subsection{Star formation rate functions}
\label{sec:SFRfunc}

Another application of our improved measures of the SFR at $z\sim4$
using both $UV$ continuum and $H\alpha$ information is for
determinations of the SFR functions \citep{Smit2012} at $z=4$-8.  The
SFR function is useful since it can be used to connect high redshift
UV-luminosity functions with H$\alpha$ and infrared (IR) based SFR
functions at $z\sim2$.

\citet{Smit2012} give a prescription to correct UV-luminosity functions for dust based on a luminosity dependent determination of the UV slope, $\beta$ \citep[e.g.][]{Bouwens2012,Bouwens2013}, a dust calibration of the form $A_{1600}=C_0+C_1\beta$ and a fixed scatter around the $\beta$-luminosity relation, $\sigma_\beta$ \citep[see Eq. 4, 7 and 8 by][]{Smit2012}. 
We assume a low metallicity stellar populations including stellar rotation, as
well as a rising star-formation history (see Table \ref{tab:fiducial}) and therefore convert the dust-corrected UV-luminosity functions to SFR functions using \citet{Kennicutt1998} relation  (Eq. \ref{Eq:Kennicutt_UV}), with no correction, since the offsets due to the star-formation history and  the higher production efficiency of ionizing photons roughly cancel each other out (see Table \ref{tab:summary}). 
At the same time, Equation \ref{Eq:Kennicutt_Ha} overpredicts SFR$_{\rm H\alpha}$
by $\sim0.2$ dex, explaining the high $\rm SFR_{\rm H\alpha}/SFR_{\rm
  UV+\beta}$ ratio.

Stepwise determinations and Schechter paramters are given in Table
\ref{tab:stepwise_sfr_SMC} and \ref{tab:schechter_sfr_SMC} respectively. We
base our SFR functions on the determination of the UV-luminosity
functions at $z\sim4$-8 by \citet{Bouwens2014} and determinations of
the color magnitude relations by \citet{Bouwens2013}. The resulting
SFR functions are shown in Figure \ref{fig:SFRfunc} in combination
with the H$\alpha$-based SFR function derived by \citet{Sobral2014},
the UV+MIR luminosity function derived by \citet{Reddy2008} and the
MIR luminosity function measured by \citet{Magnelli2011} converted to
SFR using the \citet{Kennicutt1998} relation.

On Figure \ref{fig:SFRfunc}, the SFR range above $\sim150 M_\odot\rm yr^{-1}$, equivalent to $L_{\rm bol}>10^{12}\,M_\odot$, where we might expect dust saturated sources that are missed in an UV-selected sample. Given that the $z\sim4$ and $z\sim5$ SFR functions reach beyond $L_{\rm bol}>10^{12}\,M_\odot$, we might imagine our SFR functions to be underestimated at the high end.  Our results for the SFR function are fairly similar to those recently obtained by \citet{Mashian2015}.

Instead of the SMC dust-correction, we could have assumed \citet{Meurer1999} extinction and explained the
discrepancy of UV- and H$\alpha$ based SFR estimates due to a rising star-formation history and therefore using a conversion factor 0.1 dex higher than the Kennicutt (1998) relation (Eq. \ref{Eq:Kennicutt_UV}) to match the instantaneous H$\alpha$ star-formation rates.
The resulting knee of the $z\sim4$ SFR function
would shift by $\sim+0.3$ dex.  The individual bins and Schechter
parameters derived using a \citet{Meurer1999} dust calibration are presented in appendix B
in Tables~\ref{tab:schechter_sfr}-\ref{tab:stepwise_sfr}.

Alternatively we can assume $A_{1600}=1.99\,(\beta+2.5)$, derived in \S\ref{sec:origin_dust} (see Figure \ref{fig:IRX_beta}) to bring H$\alpha$ and UV based SFRs into agreement when assuming constant star-formation histories. Implementing this assumption into our SFR functions would imply a big shift of $\sim+0.5$ dex shift in the high-end of SFR function at $z\sim4$, resulting in similar SFR functions at $z\sim2$ and $z\sim4$ (see Figure \ref{fig:SFRfunc}). As a consequence the total star formation rate density does not decline after $z\sim2$, but plateau out to $z\sim4$ and decline at $z\gtrsim5$.


The systematic uncertainty in the present SFR function will be
alleviated when more observations with new generation sub-mm
facilities such as ALMA become available over the next few years.

\section{Summary}
\label{sec:summary}

In this paper we make use of a large sample of galaxies with
spectroscopic redshifts between $z=3.8$-5.0, where H$\alpha$ can be
inferred from the excess in the $3.6\,\micron$ \textit{Spitzer}/IRAC
band, as well as a photometric sample in the same redshift range. As
in previous studies \citep[e.g.][]{Shim2011,Stark2013} we find a
typical rest-frame H$\alpha$ EW of $\sim$ 400 {\AA} for a
spectroscopic $z=3.8$-5.0 sample.  In addition, we also conduct a
systematic investigation of the H$\alpha$ EWs in pure
photometric-redshift-selected $z\sim3.8$-5.0 sample and find similar
results for both samples \citep[see also][]{Rasappu2015,Marmol2015}.  While we find
no strong dependence of the H$\alpha$ EWs on UV luminosity, UV
slope, half-light radius or S\'{e}rsic index, we do however find a
clear relation between EW$_0$(H$\alpha$+[\ion{N}{2}]+[\ion{S}{2}]) and
mass-to-light ratio, $M_\ast/L_{\rm UV}$ (Figure \ref{fig:EWs}).

We explore the use of the inferred H$\alpha$ fluxes to derive star
formation rates for galaxies in our samples. We compare these
H$\alpha$-based SFRs with UV-based SFRs using the \citet{Meurer1999}
relation and find a strong correlation between the two
estimates. However, even when we assume similar extinction towards
nebular regions and stellar populations, i.e. $A_{V,\rm stars}= A_{V,
  \rm gas}$, we still find a small systematic offset $\sim0.10-0.16$ dex in
the $\rm SFR_{\rm H\alpha}/SFR_{\rm UV+\beta}$ ratios of both our
samples.

In this paper, we consider the impact of the assumed  dust law, SFH and the shape of the
ionizing spectrum on the $\rm SFR_{\rm
  H\alpha}/SFR_{\rm UV+\beta}$ ratio.  Here we provide a summary of
our conclusions:

\begin{itemize}

\item \textit{Dust law:} The largest uncertainty in our UV-based SFRs is the dust law. While one issue is the reddening law, another issue is a potential evolution in the intrinsic color of galaxies (prior to the impact of dust reddening). In particular, galaxies with low metallicities and young ages can have bluer intrinsic UV slopes than those of the galaxies in the \citet{Meurer1999} calibration, which would result in an underestimate of the dust content in our galaxies. We investigate the typical intrinsic UV-continuum slope needed to explain the offsets in H$\alpha$ and UV based SFRs and find $A_{1600}=1.99\,(\beta+2.5)$. This dust correction is in agreement with FIR stacking measurements \citep{Coppin2015}, but differs quite strongly from recent ALMA measurements \citep[e.g.][]{Capak2015,Aravena2016,Bouwens2016,Dunlop2016}. Assuming an SMC dust-law such as favoured by \citet{Capak2015} and \citet{Bouwens2016} the $\rm SFR_{\rm H\alpha}/SFR_{\rm UV+\beta}$ ratio would be offset by $\sim0.2$ dex (assuming $A_{V,\rm stars}= A_{V, \rm gas}$). 

\item \textit{Bursty star-formation history:} A natural consequence of bursty star formation histories is to produce high H$\alpha$ EWs and high $\rm SFR_{\rm H\alpha}/SFR_{\rm UV+\beta}$ ratios for short ($\sim5$ Myr) time periods. However, we find comparable $\rm SFR_{\rm H\alpha}/SFR_{\rm UV+\beta}$ for both our spectroscopic-redshift and photometric-redshift sample and even a mass limited photometric subsample. We use a Monte Carlo simulation to compare the expected sSFRs from H$\alpha$ and UV indicators with our samples. We find that a sample of galaxies with typical burst masses of  $M_{\rm burst}\sim10^8 M_\odot$ and burst intervals of $\rm dt_{burst}\sim5-10 Myr$ can produce a $\sim0.1$ dex offset in  $\rm SFR_{\rm H\alpha}/SFR_{\rm UV+\beta}$. However, we also find a low  $\rm sSFR_{\rm H\alpha}$ tail in our simulated distribution that is $\sim2\times$ larger than we find in our observed sample, which argues against significantly bursty star-formation histories.

\item \textit{Rising star-formation history:} Rising star formation histories create an offset in the $\rm SFR_{\rm H\alpha}/SFR_{\rm UV+\beta}$ ratio, due to the different timescales of star formation probed by H$\alpha$ ($\sim10$ Myr) and UV ($\sim100$ Myr) SFR indicators. We estimate this offsets the $\rm SFR_{\rm H\alpha}/SFR_{\rm UV+\beta}$ ratio by $\sim0.1$ dex, using prescriptions given in \citet{Reddy2012}. 

\item \textit{Production Efficiency for Ionizing Photons / Ionizing Spectrum:} The shape of the ionizing radiation field for low metallicity stellar populations is currently poorly constrained, which is unfortunate since this can have a significant impact on the nebular emission of high redshift galaxies \citep[e.g.][]{Kewley2013}. We investigate two sets of models that include the effects of stellar rotation \citep{Leitherer2014} and effects of massive binary star systems \citep{Eldridge2012} to low metallicity. We find that these models can generate an observed offset in the $\rm SFR_{\rm H\alpha}/SFR_{\rm UV+\beta}$ ratio of $\sim0.1-0.3$ dex.
We argue that, if high-redshift UV-selected galaxies prefer an SMC type dust-law, some source of ionizing photons is likely required on top of e.g. a rising star-formation history to explain the strength of the H$\alpha$ emission lines we observe.

\end{itemize}


We find that there are two flavors of physical models that appear
plausible on the basis of the observations we consider
(Table~\ref{tab:fiducial}): (1) the first invoking a
\citet{Meurer1999} dust calibration with a $A_{V,\rm stars}= A_{V, \rm
  gas}$ and correcting up $UV$-based, time-averaged SFR estimates by
$\sim$0.1 dex to better match the instantaneous SFRs and (2) the second
invoking a SMC dust law with a $A_{V,\rm stars}= A_{V, \rm gas}$ and
supposing that $z\sim4$ galaxies are more efficient at producing
ionizing photons than in standard stellar population models
\citep[see][for a discussion of the impact this may have of galaxies
  as being capable of reionizing the universe]{Bouwens2015,Stark2015,Stark2016}.

We adopt the latter flavor of physical model as our fiducial one (and
include some results from the former model in appendix B).  We use
this model to construct the main sequence of star forming galaxies
from our H$\alpha$-based SFRs (\S\ref{sec:mainseq}).  We find that,
when taking into account the incompleteness at the faint end of our
selection, the slope is broadly consistent with the unity low-mass
slope found by \citet{Whitaker2014} at $z\sim0.5-2.5$. Furthermore,
while we find an intrinsic scatter of $\sigma\sim0.4$ and a
normalization of the main sequence of $\rm \log_{10} sSFR_{\rm
  H\alpha}/\rm Gyr^{-1}=0.47\pm0.06$, in reasonable agreement with
   recent determinations from SED fitting
\citep[e.g.][]{Gonzalez2014,Duncan2014,Salmon2015}.

In \S\ref{sec:SFRfunc}, we follow the \citet{Smit2012} procedure to
infer SFR functions at $z\sim4$-8 from the UV luminosity functions
derived by \citet{Bouwens2014}, and the UV-continuum slopes derived by
\citet{Bouwens2013}.  Consistent with our fiducial approach, we use
a SMC-type dust calibration and we assume the \citet{Kennicutt1998} 
conversion factor between UV and SFR, given that the effects of an 
increased production efficiency of ionizing photons and a
 rising star-formation history roughly cancel each other 
 out \citep[see also][]{Mashian2015}.  The $z=4$-8 SFR
functions for a \citet{Meurer1999} dust correction are presented in appendix B.

We conclude that systematic use of H$\alpha$ star-formation rates inferred  from \textit{Spitzer}/IRAC photometry provides an exciting opportunity to unravel fundamental properties of the high-redshift galaxy population in advance of JWST. 

\acknowledgments
We thank Selma de Mink, Carlos Frenk, Ylva G\"otberg, John Lucey and Tom Theuns for interesting conversations. This work was supported by the Leverhulme Trust.

\begin{table}
\begin{threeparttable}
\centering
\caption{Parameters of the SFR$_{\rm H\alpha}-M_\ast$ sequence}
\begin{tabular}{lc} 
\hline \hline 
$d{\rm\,log_{10}(SFR)}/d\,{\rm log_{10}}(M_\ast)$ & 0.78$\pm$0.23\\
$\rm log_{10}(SFR_{\rm H\alpha})_{{M_\ast=10^{10}\,M_\odot }}$ [$M_\odot\rm /yr^{-1}$] & 1.51$\pm$0.07\\
$\sigma_{\rm intrinsic}$ & 0.36$\pm$0.06 \\
\hline 
\end{tabular} 
\label{tab:sfr_mass}
\begin{tablenotes}
\item Based on the Bayesian linear regression of all sources in our combined photometric and spectroscopic sample (with sources in both samples counted once) with $M_\ast>10^{9.8}M_\odot$.  
\end{tablenotes}
\end{threeparttable}
\vspace{1cm}
\end{table}

\begin{table}[h]
\begin{threeparttable}
\centering
\caption{Schechter parameters of the SFR functions: SMC dust correction}
\begin{tabular}{cccc}
\centering
$<z>$ & log$_{10}$ $\frac{\rm SFR^\ast}{M_\odot \rm yr^{-1}}$ & $\phi_{\rm SFR}^\ast$ (10$^{-3}$ Mpc$^{-3}$) & $\alpha_{\rm SFR}$ \\ 
\hline \hline 
3.8&1.41$\pm$0.04& 1.76$^{+0.30}_{-0.26}$ &$-$1.57$\pm$0.06\\
4.9&1.53$\pm$0.06& 0.65$^{+0.16}_{-0.12}$ &$-$1.66$\pm$0.08\\
5.9&1.42$\pm$0.10& 0.41$^{+0.18}_{-0.13}$ &$-$1.72$\pm$0.16\\
6.8&1.37$\pm$0.14& 0.27$^{+0.18}_{-0.10}$ &$-$1.82$\pm$0.23\\
7.9&1.19$\pm$0.24& 0.18$^{+0.20}_{-0.10}$ &$-$1.91$\pm$0.41\\
\hline
\end{tabular} 
\label{tab:schechter_sfr_SMC}
\begin{tablenotes}
\item These Schechter parameters are obtained following the procedure described \citet{Smit2012}. We assume an SMC dust correction (Eq. \ref{Eq:SMCdust}) and adopt the linear relation between the UV-continuum slope $\beta$ and UV luminosity found by \citet{Bouwens2013}, see \S \ref{sec:SFRfunc} and we assume the \citet{Kennicutt1998} conversion from UV to SFR (Eq. \ref{Eq:Kennicutt_UV}).
\end{tablenotes}
\end{threeparttable}
\end{table}

\begin{table*}
\begin{center}
\begin{threeparttable}
\caption{Stepwise determinations of the SFR function at $z\sim4$, $z\sim5$, $z\sim6$, $z\sim7$ and $z\sim8$: SMC dust correction }
\begin{tabular}{lrclr}

\hline \hline 
log$_{10}\,\rm{SFR}\,(M_\odot\rm{yr^{-1}})$&$\rm\phi_{SFR}\,(Mpc^{-3}dex^{-1})$ &$\,\,\,\,$ & log$_{10}\,\rm{SFR}\,(M_\odot\rm{yr^{-1}})$&$\rm\phi_{SFR}\,(Mpc^{-3}dex^{-1})$\\ 
\cline{1-2}
\cline{4-5}
\multicolumn{2}{c}{$z\sim4$} & & \multicolumn{2}{c}{$z\sim6$}\\
\cline{1-2}
\cline{4-5}
$-$0.76$\pm$ 0.02& 0.058688$\pm$0.018393 & & $-$0.47$\pm$ 0.02& 0.032668$\pm$0.010059\\
$-$0.33$\pm$ 0.02& 0.056971$\pm$0.008112 & & $-$0.05$\pm$ 0.03& 0.015293$\pm$0.003159\\
$-$0.01$\pm$ 0.02& 0.024028$\pm$0.005114 & &  0.40$\pm$ 0.04& 0.004192$\pm$0.000706\\
 0.21$\pm$ 0.03& 0.015637$\pm$0.002376 & &  0.75$\pm$ 0.05& 0.002675$\pm$0.000294\\
 0.43$\pm$ 0.03& 0.012133$\pm$0.001011 & &  0.98$\pm$ 0.05& 0.001474$\pm$0.000175\\
 0.65$\pm$ 0.03& 0.006738$\pm$0.000576 & &  1.22$\pm$ 0.05& 0.000670$\pm$0.000086\\
 0.87$\pm$ 0.03& 0.005570$\pm$0.000416 & &  1.46$\pm$ 0.06& 0.000366$\pm$0.000052\\
 1.10$\pm$ 0.03& 0.003808$\pm$0.000254 & &  1.70$\pm$ 0.06& 0.000110$\pm$0.000025\\
 1.32$\pm$ 0.03& 0.001519$\pm$0.000141 & &  1.94$\pm$ 0.06& 0.000031$\pm$0.000012\\
 1.54$\pm$ 0.03& 0.000879$\pm$0.000089 & &  2.18$\pm$ 0.06& 0.000004$\pm$0.000004\\
  \cline{4-5}
 1.77$\pm$ 0.03& 0.000299$\pm$0.000051 & &  \multicolumn{2}{c}{$z\sim7$}\\
 \cline{4-5}
 1.99$\pm$ 0.03& 0.000034$\pm$0.000020 & &  $-$0.42$\pm$ 0.03& 0.020132$\pm$0.006963\\
 2.21$\pm$ 0.03& 0.000007$\pm$0.000009 & &   0.01$\pm$ 0.05& 0.013461$\pm$0.003365\\
 \cline{1-2}
 \multicolumn{2}{c}{$z\sim5$}        & &       0.34$\pm$ 0.06& 0.003842$\pm$0.001070\\
 \cline{1-2}
$-$0.61$\pm$ 0.02& 0.057283$\pm$0.016809 & &     0.57$\pm$ 0.07& 0.001984$\pm$0.000387\\
$-$0.18$\pm$ 0.02& 0.019776$\pm$0.004047 & &     0.80$\pm$ 0.08& 0.001410$\pm$0.000216\\
 0.26$\pm$ 0.03& 0.010087$\pm$0.001221 & &     1.03$\pm$ 0.08& 0.000659$\pm$0.000130\\
 0.60$\pm$ 0.03& 0.004661$\pm$0.000382 & &     1.27$\pm$ 0.08& 0.000408$\pm$0.000072\\
 0.82$\pm$ 0.03& 0.002949$\pm$0.000209 & &     1.51$\pm$ 0.09& 0.000101$\pm$0.000032\\
 1.05$\pm$ 0.03& 0.002274$\pm$0.000148 & &     1.75$\pm$ 0.09& 0.000069$\pm$0.000019\\
 1.27$\pm$ 0.03& 0.001489$\pm$0.000101 & &     1.98$\pm$ 0.09& 0.000002$\pm$0.000004\\
    \cline{4-5}
 1.50$\pm$ 0.03& 0.000582$\pm$0.000055 & &   \multicolumn{2}{c}{$z\sim8$}\\ 
    \cline{4-5}
 1.73$\pm$ 0.03& 0.000222$\pm$0.000031 & &   $-$0.12$\pm$ 0.12& 0.006478$\pm$0.002459\\
 1.96$\pm$ 0.03& 0.000074$\pm$0.000018 & &    0.31$\pm$ 0.15& 0.002470$\pm$0.000792\\
 2.18$\pm$ 0.03& 0.000013$\pm$0.000007 & &    0.63$\pm$ 0.16& 0.001222$\pm$0.000518\\
 2.41$\pm$ 0.04& 0.000004$\pm$0.000004 & &    0.85$\pm$ 0.17& 0.000750$\pm$0.000236\\
                                 & & &        1.08$\pm$ 0.18& 0.000135$\pm$0.000058\\
                                 & & &        1.30$\pm$ 0.18& 0.000129$\pm$0.000033\\
                                 & & &        1.52$\pm$ 0.19& 0.000029$\pm$0.000011\\
                                 & & &        1.75$\pm$ 0.19& 0.000011$\pm$0.000007\\
 \hline

\end{tabular} 
\label{tab:stepwise_sfr_SMC}

These Schechter parameters are obtained following the procedure described \citet{Smit2012}. We assume an SMC dust correction (Eq. \ref{Eq:SMCdust}) and adopt the linear relation between the UV-continuum slope $\beta$ and UV luminosity found by \citet{Bouwens2013}, see \S \ref{sec:SFRfunc} and we assume the \citet{Kennicutt1998} conversion from UV to SFR (Eq. \ref{Eq:Kennicutt_UV}).
\end{threeparttable}
\end{center}
 \end{table*}

\bibliographystyle{plainnat}

\appendix

\section{A. Testing Our Method to Extract Line Fluxes from Photometric Observations}
\label{appendixtest}

To validate our method of inferring H$\alpha$ using the \textit{Spitzer}/IRAC photometry, we perform two tests at $z<3.8$. 
First, we assemble a small sample of lower redshift galaxies with emission line measurements obtained directly from spectroscopy and compare these measurements with the values inferred from the flux offset in the contaminated band from the continuum SED. We use a sample of  line emitters selected from the 3D-HST grism survey presented by \citet{Momcheva2016} at $z=1.3-1.5$ where H$\alpha$, [\ion{N}{2}] and [\ion{S}{2}] contaminate the $H_{160}$ band. From the 3D-HST survey we select sources with a $>5\sigma$ detection of the H$\alpha$+[\ion{N}{2}] blended emission lines, a $>5\sigma$ detection in the K-band (the adjacent band to the $H_{160}$ band used to derive emission lines) and a $V_{606}$-band total magnitude brighter than 26.5 mag (the $V_{606}$ band at $z=1.3-1.5$ corresponds to roughly the same rest-frame wavelength as the $H_{160}$ band at $z=3.8-5.0$).

\begin{figure*}
\centering
\includegraphics[width=.48\columnwidth,trim=100mm 175mm 25mm 35mm,clip] {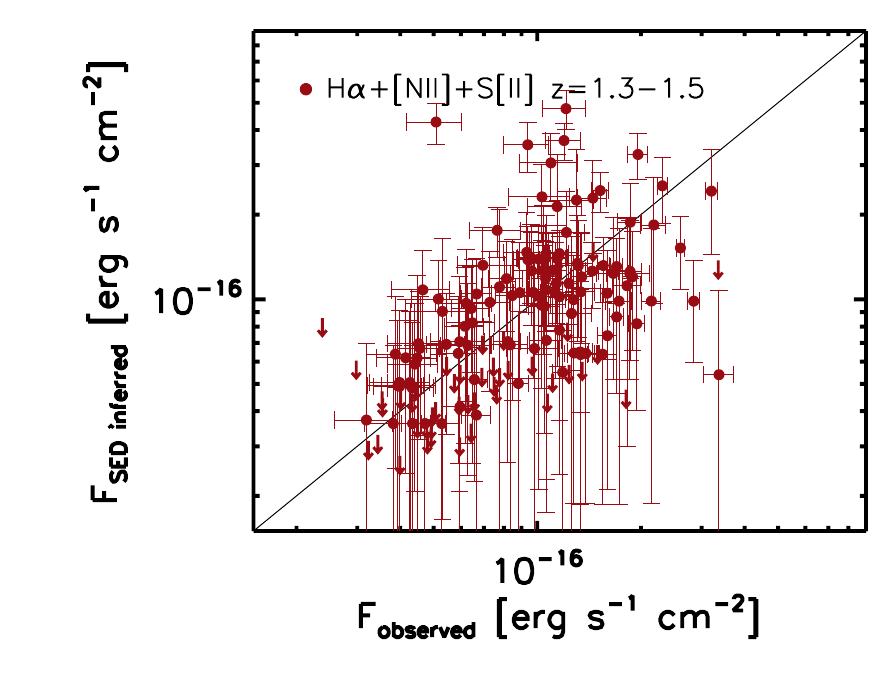} 
\includegraphics[width=.48\columnwidth,trim=100mm 175mm 25mm 35mm,clip] {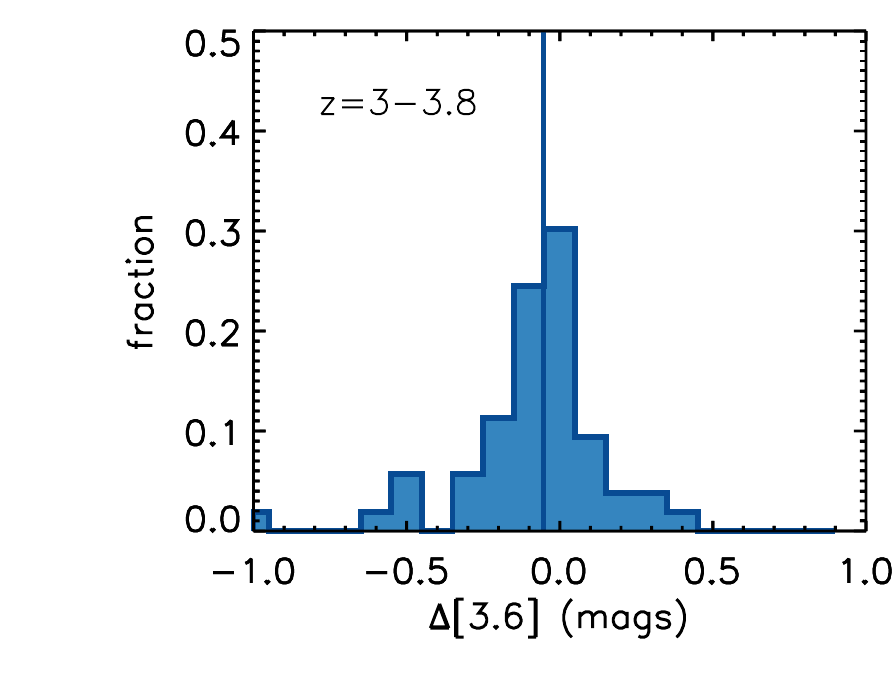} 
\caption{\textit{Left panel:} A test of our inferred emission line method using galaxies at $z=1.3-1.5$ \citep{Momcheva2016} where the the H$\alpha$+[\ion{N}{2}]+[\ion{S}{2}] lines contaminate the $H_{160}$ band. The line fluxes inferred from the photometry are  -0.01 dex offset in the median source from the spectroscopic measurements. \textit{Right panel:} The 3.6$\micron$ excess flux measurements of a spectroscopic sample at $z=3.0-3.8$, where the 3.6$\micron$ band is unaffected by strong emission lines and should therefore show no offsets from the predicted continuum flux. The sample is -0.05 magnitude or -0.02 dex below the predicted continuum flux in the median source.  } 
\label{fig:tests}
\vspace*{3mm}
\end{figure*}

For this sample of 160 sources we use the same photometry as described in \S\ref{sec:spec_sample} and we perform the same steps in deriving the SED inferred line fluxes as described in \S\ref{sec:prop}. We do not include photometry bands contaminated with strong emission lines and we assume the same fixed line ratios between H$\alpha$, [\ion{N}{2}] and [\ion{S}{2}] as in our main sample and correct the H$\alpha$+[\ion{N}{2}] measurement by 10\% to account for the flux of [\ion{S}{2}] (the grism measurements of this line are highly uncertain or not detected at all). We present the results of our inferred line fluxes in the left panel of Figure \ref{fig:tests}. We find a median difference of $-0.01\pm0.02$ dex (uncertainty obtained through bootstrapping) between the line fluxes measured with spectroscopy and those inferred from the photometry.

For our second test we collected a sample of spectroscopically confirmed galaxies between $z=3.0-3.8$, using the same public redshift catalogs of \citet{Balestra2010,Vanzella2005,Vanzella2006,Vanzella2008} over the GOODS-S field as used for our spectroscopic sample described in \S\ref{sec:spec_sample}. For this sample we will test the the same methods and photometry that we have used in our paper for our spectroscopic $z=3.8-5.0$ galaxy sample, by deriving the offset in the $3.6\micron$ band flux from the predicted continuum flux, which given the lack of line contamination at $z=3.0-3.8$ should be approximately zero. An important difference for the   $z=3.0-3.8$  samples is that 70\% and 90\% of the galaxies have contaminated photometry from the  [\ion{O}{3}] and  [\ion{O}{2}] emission lines respectively, while for the $z=3.8-5.0$ sample only 43\% of all the galaxies can have affected photometry due to the flux of  the [\ion{O}{2}] emission lines. Since the photometry of the $z=3.0-3.8$ sample is significantly affected by emission lines, we exclude the $JH_{140}, H_{160}$ and $K_{s}$-band from the SED fitting, but otherwise keep all steps identical to the procedures described in \S\ref{sec:spec_sample}-\ref{sec:prop}.

The results of the offsets between $3.6\micron$ band flux from the predicted continuum flux is presented in the right panel of figure \ref{fig:tests}. We find a small median offset of $-0.05\pm0.01$ magnitude or $-0.02\pm0.01$ dex (uncertainty obtained through bootstrapping) in our $z=3.0-3.8$ spectroscopic galaxy sample, with a modest standard deviation of 0.23 magnitude or 0.09 dex.

Both these test results indicate that there are no obvious systematics present in our photometry that would cause us to \textit{overestimate} the inferred H$\alpha$ flux, given our method.  We remark, however, that it is possible that we might be underestimating the derived H$\alpha$ luminosity.  This underestimate, if real, would strengthen our conclusion that there are small systematic differences between H$\alpha$ and UV-based SFRs given standard low-redshift calibrations.

\section{B. SFR functions assuming a SMC dust correction}
\label{appendix}

 In this appendix
we present SFR functions as described in \S\ref{sec:SFRfunc} and using
the \citet{Meurer1999} dust correction in Equation \ref{Eq:Meurer}.  These results
correspond to the second fiducial model presented in
Table~\ref{tab:fiducial}.

In deriving the SFR functions, we dust correct the $UV$ continuum
light using a SMC extinction law and do not make any further
corrections.  In addition, we convert the $UV$ luminosities into SFR
using Equation \ref{Eq:Kennicutt_UV} (Equation \ref{Eq:Kennicutt_Ha}
over predicts the H$\alpha$ SFR by $\sim0.2$ dex).  In doing so, we
rely on the conclusions from \S
\ref{sec:origin_dust}-\ref{sec:origin_ionrad} where we find that a
combination of rising star-formation histories and low metallicity
stellar population models including stellar rotation can bring
H$\alpha$ and UV-based SFR estimates into good agreement.

\begin{table}
\begin{threeparttable}
\centering
\caption{Schechter parameters of the SFR functions: M99 dust correction}
\begin{tabular}{cccc}
\centering
$<z>$ & log$_{10}$ $\frac{\rm SFR^\ast}{M_\odot \rm yr^{-1}}$ & $\phi_{\rm SFR}^\ast$ (10$^{-3}$ Mpc$^{-3}$) & $\alpha_{\rm SFR}$ \\ 
\hline \hline 
3.8&1.74$\pm$0.07& 1.62$^{+0.28}_{-0.24}$ &$-$1.53$\pm$0.05\\
4.9&1.87$\pm$0.08& 0.59$^{+0.14}_{-0.11}$ &$-$1.60$\pm$0.07\\
5.9&1.75$\pm$0.14& 0.36$^{+0.16}_{-0.12}$ &$-$1.63$\pm$0.14\\
6.8&1.68$\pm$0.20& 0.23$^{+0.16}_{-0.09}$ &$-$1.73$\pm$0.21\\
7.9&1.43$\pm$0.38& 0.16$^{+0.18}_{-0.09}$ &$-$1.85$\pm$0.38\\
\hline
\end{tabular} 
\label{tab:schechter_sfr}
\begin{tablenotes}
\item These Schechter parameters are obtained following the procedure described \citet{Smit2012}. We assume a \citet{Meurer1999} dust correction and adopt the linear relation between the UV-continuum slope $\beta$ and UV luminosity found by \citet{Bouwens2013}, see \S \ref{sec:SFRfunc}.  Moreover, we assume a $0.07$ dex increase on the \citet{Kennicutt1998} conversion from UV to SFR to better match the instantaneous H$\alpha$ star-formation rates (see \S\ref{sec:SFRfunc}).
\end{tablenotes}
\end{threeparttable}
\end{table}

\begin{table*}
\begin{center}
\begin{threeparttable}
\caption{Stepwise determinations of the SFR function at $z\sim4$, $z\sim5$, $z\sim6$, $z\sim7$ and $z\sim8$: M99 dust correction }
\begin{tabular}{lrclr}

\hline \hline 
log$_{10}\,\rm{SFR}\,(M_\odot\rm{yr^{-1}})$&$\rm\phi_{SFR}\,(Mpc^{-3}dex^{-1})$ &$\,\,\,\,$ & log$_{10}\,\rm{SFR}\,(M_\odot\rm{yr^{-1}})$&$\rm\phi_{SFR}\,(Mpc^{-3}dex^{-1})$\\ 
\cline{1-2}
\cline{4-5}
\multicolumn{2}{c}{$z\sim4$} & & \multicolumn{2}{c}{$z\sim6$}\\
\cline{1-2}
\cline{4-5}
-0.63$\pm$ 0.04& 0.055128$\pm$0.017277 & & -0.39$\pm$ 0.03& 0.031227$\pm$0.009615\\
-0.17$\pm$ 0.04& 0.053183$\pm$0.007573 & &  0.07$\pm$ 0.06& 0.014083$\pm$0.002909\\
 0.18$\pm$ 0.05& 0.022292$\pm$0.004744 & &  0.56$\pm$ 0.08& 0.003761$\pm$0.000633\\
 0.42$\pm$ 0.05& 0.014455$\pm$0.002196 & &  0.95$\pm$ 0.09& 0.002365$\pm$0.000260\\
 0.66$\pm$ 0.05& 0.011193$\pm$0.000933 & &  1.22$\pm$ 0.10& 0.001296$\pm$0.000154\\
 0.90$\pm$ 0.05& 0.006207$\pm$0.000530 & &  1.49$\pm$ 0.10& 0.000588$\pm$0.000075\\
 1.14$\pm$ 0.05& 0.005127$\pm$0.000383 & &  1.77$\pm$ 0.10& 0.000321$\pm$0.000046\\
 1.38$\pm$ 0.05& 0.003503$\pm$0.000233 & &  2.04$\pm$ 0.10& 0.000096$\pm$0.000022\\
 1.62$\pm$ 0.05& 0.001397$\pm$0.000130 & &  2.32$\pm$ 0.10& 0.000027$\pm$0.000011\\
 1.86$\pm$ 0.05& 0.000809$\pm$0.000082 & &  2.59$\pm$ 0.10& 0.000004$\pm$0.000004\\
  \cline{4-5}
 2.11$\pm$ 0.06& 0.000275$\pm$0.000047 & &  \multicolumn{2}{c}{$z\sim7$}\\
 \cline{4-5}
 2.35$\pm$ 0.06& 0.000031$\pm$0.000018  & &  -0.32$\pm$ 0.06& 0.019064$\pm$0.006594\\
 2.59$\pm$ 0.06& 0.000006$\pm$0.000008  & &  0.13$\pm$ 0.10& 0.012464$\pm$0.003116\\
 \cline{1-2}
 \multicolumn{2}{c}{$z\sim5$}        & &     0.50$\pm$ 0.13& 0.003480$\pm$0.000969\\
 \cline{1-2}
-0.49$\pm$ 0.04& 0.053846$\pm$0.015801 & &      0.75$\pm$ 0.14& 0.001773$\pm$0.000346\\
-0.03$\pm$ 0.05& 0.018326$\pm$0.003750 & &      1.01$\pm$ 0.15& 0.001252$\pm$0.000191\\
 0.45$\pm$ 0.05& 0.009254$\pm$0.001120 & &      1.27$\pm$ 0.15& 0.000582$\pm$0.000115\\
 0.82$\pm$ 0.06& 0.004252$\pm$0.000349 & &      1.54$\pm$ 0.16& 0.000360$\pm$0.000063\\
 1.06$\pm$ 0.06& 0.002683$\pm$0.000190 & &      1.81$\pm$ 0.16& 0.000089$\pm$0.000028\\
 1.31$\pm$ 0.06& 0.002066$\pm$0.000135 & &      2.08$\pm$ 0.16& 0.000061$\pm$0.000017\\
 1.56$\pm$ 0.06& 0.001352$\pm$0.000092 & &      2.35$\pm$ 0.16& 0.000002$\pm$0.000004\\
    \cline{4-5}
 1.81$\pm$ 0.06& 0.000529$\pm$0.000050 & &   \multicolumn{2}{c}{$z\sim8$}\\ 
    \cline{4-5}
 2.06$\pm$ 0.06& 0.000201$\pm$0.000028& &     -0.01$\pm$ 0.24& 0.006130$\pm$0.002327\\
 2.31$\pm$ 0.06& 0.000068$\pm$0.000016 & &     0.44$\pm$ 0.28& 0.002311$\pm$0.000741\\
 2.56$\pm$ 0.06& 0.000012$\pm$0.000006 & &     0.79$\pm$ 0.31& 0.001131$\pm$0.000479\\
 2.82$\pm$ 0.06& 0.000004$\pm$0.000004 & &     1.03$\pm$ 0.32& 0.000689$\pm$0.000216\\
                                 & & &          1.27$\pm$ 0.33& 0.000123$\pm$0.000053\\
                                 & & &          1.52$\pm$ 0.34& 0.000118$\pm$0.000031\\
                                 & & &          1.77$\pm$ 0.34& 0.000026$\pm$0.000010\\
                                 & & &          2.01$\pm$ 0.34& 0.000010$\pm$0.000006\\
  
 \hline

\end{tabular} 
\label{tab:stepwise_sfr}

These SFR functions are obtained following the procedure described \citet{Smit2012}. We dust correct the stepwise UV LFs by \citet{Bouwens2014} using the \citet{Meurer1999} IRX-$\beta$ relationship.  We adopt the linear relation between the UV-continuum slope $\beta$ and UV luminosity found by \citet{Bouwens2013}, see \S \ref{sec:SFRfunc}. Moreover, we assume a $0.07$ dex increase on the \citet{Kennicutt1998} conversion from UV to SFR to better match the instantaneous H$\alpha$ star-formation rates (see \S\ref{sec:SFRfunc}).
\end{threeparttable}
\end{center}
 \end{table*}


\end{document}